\documentclass[a4paper]{jpconf}
\usepackage{graphicx}

\usepackage{amssymb}
\usepackage{color}%

\usepackage{chngcntr}

\begin{document}

\title{Alternative Formulation of the Induced Surface and Curvature 
Tensions Approach}


\author{Kyrill A. Bugaev}

\address{  Bogolyubov Institute for Theoretical Physics,
Metrologichna str. 14$^B$, Kyiv 03680, Ukraine}

\address{ Department of Physics, Taras Shevchenko National University of Kyiv, 03022 Kyiv, Ukraine}

\ead{bugaev@th.physik.uni-frankfurt.de}

\begin{abstract}
We develop a novel method to analyze the excluded volume of the multicomponent mixtures of
classical hard spheres in the grand canonical ensemble. 
The method is based on the Laplace-Fourier transform technique and allows one to 
account for the fluctuations of the particle number density  for  the induced surface  and curvature tensions equation of state. As a result one can go beyond the Van der Waals approximation by obtaining the suppression of the  induced surface  and curvature tensions coefficients at moderate and 
high packing fractions. In contrast to the standard induced surface  and curvature tensions equation of state the suppression of these coefficients is not  an exponential one, but a power-like one. The obtained alternative equation of state is further generalized to account for higher virial coefficients.
This result is straightforwardly generalized to the case of quantum statistics.
\end{abstract}

\section{Introduction}
Investigation of the strongly interacting matter phase diagram 
requires  the development of realistic equation of state (EoS) which can  reliably  model  the mixtures
of hadrons, light nuclei and bags of quark-gluon plasma. Yet this is an  unresolved task, since in the 
collisions of heavy ions at high energies the number of hadrons  is not conserved and, therefore, instead of canonical treatment which is popular in statistical mechanics of classical and 
non-relativistic systems one has to use the grand canonical ensemble. Since this task is not resolved yet,  the existing 
exactly solvable EoS to model   the strongly interacting matter phase transitions \cite{Bugaev07,Bugaev+Reuter,BugaevFWM,Alexei12a,Alexei12b,Bugaev13} cannot presently  provide  the quantitative description of  its phases. 

The main problem is that to model the strongly interacting matter phase transitions  one has to go beyond the Van der Waals (VdW) approximation  and  to consider  the third, fourth and higher virial 
coefficients for the constituents which  a priori have different sizes. Note that even for the hard spheres of different radii this problem is not resolved yet \cite{Simple_Liquids,Mulero}.   In addition the problem gets harder due to the fact that  the quark-gluon bags can have any volume above some minimal value   \cite{Bugaev07,Bugaev+Reuter,BugaevFWM,Alexei12a,Alexei12b,Bugaev13} and, hence, it is necessary to  develop an approach which  is able to accurately  account  for an  infinite number of  different  hard-core radii of such bags. 

Fortunately,  the situation is getting better after an  invention of the induce surface tension (IST) EoS \cite{IST0, IST1,IST1b,IST1c,IST2,IST2a,Veta19,Veta20,IST2b,IST2020a,IST2020b,IST2020c}. The IST EoS was invented and successfully applied to the mixture of nuclear fragments of all sizes to describe the nuclear liquid-gas phase transition with a compressible  nuclear liquid phase \cite{IST0}. Then IST EoS  provided the best of existed description of hadronic multiplicities  measured in heavy ion collisions from the center-of-mass energy $\sqrt{s_{NN}} =2.7 $ GeV (AGS BNL data)  to  $\sqrt{s_{NN}} =2.76 $ TeV (ALICE CERN data) for  the same set of hard-core radii with the total quality $\chi^2/dof =1.13$ \cite{ IST1,IST1b,IST2}.  
In addition the IST EoS with quantum statistics \cite{IST2a} was able to successfully reproduce 
the basic properties of symmetric nuclear matter and the so called proton flow constraint 
\cite{ProtonFlow} having  four adjustable parameters only \cite{IST1c}. This is highly nontrivial
result, since these four parameters allow one to simultaneously reproduce the properties of normal nuclear matter (4 conditions) and the proton flow constraint 
(8 conditions) which in total make 12 conditions. 
{ The success of one-component IST EoS for symmetric nuclear matter \cite{IST2a} initiated a direction of the
successful applications of  such EoS to  model  the neutron stars properties \cite{Veta19,Veta20}.}

Very recently the IST EoS was successfully extended to  reproduce the classical  second virial coefficients of light nuclei \cite{IST2b,IST2020a,IST2020b,IST2020c} and  such an approach allowed us for the first time to  simultaneously describe the hadron and light (anti-, hyper-) nuclei multiplicities measured in heavy ion collisions by the STAR BNL  Collaboration at $\sqrt{s_{NN}} =200 $ GeV and by the ALICE CERN Collaboration at $\sqrt{s_{NN}} =2760$ GeV with very high quality of data description $\chi^2/dof =1.07$.  Note that compared to other versions of the hadron resonance gas model with the hard-core repulsion  \cite{HRGM1,HRGM2,HRGM3} only the IST EoS is  perfectly suited to take into account the second virial coefficients of such composite objects as light (anti-, hyper-) nuclei. Moreover, the IST EoS has two  great advantages over the other 
versions of the hadron resonance gas model:  first,  it allows one to easily  take into account not only the second virial coefficients, but also the third and the fourth ones, and, second,  the number of  equations to be solved is two and it does not depend on the number of considered different hard-core radii. 

However, despite these advantages the IST EoS is valid up to packing fractions $\eta = \sum_k V_k \rho_k= 0.2-0.22$  \cite{ IST1,IST2}, where $V_k = \frac{4}{3} \pi R_k^3$ is the eigenvolume of the $k$-th sort of particles which have  the particle number density $\rho_k$ and the hard-core radius $R_k$ (for a comparison we mention that  the VdW EoS with hard-core repulsion is valid for $\eta \le 0.1-0.11$  \cite{ IST1,IST2}).
Therefore,  recently in the IST EoS the curvature term was included and  the  induce surface and curvature  tensions (ISCT) EoS was worked out in Refs.  \cite{ISCT1,ISCT2}.  In Ref.  \cite{ISCT2}
 the ISCT EoS was thoroughly compared to the two-component mixtures of classical  hard spheres and classical  hard discs  and it was found that the ISCT EoS is able to very accurately 
  reproduce  the multi-component version of the famous 
Carnahan-Starling (CS) EoS of  hard spheres, known as the  the Mansoori-Carnahan-Starling-Leland  (MCSL) EoS \cite{MCSL} (for hard spheres),  up to  $\eta \simeq 0.45$ and  the EoS of hard discs up to $\eta \simeq 0.7$. In other words, the ISCT EoS is applicable to entire gaseous phase till the 
phase transition to the  solid state \cite{Simple_Liquids, Mulero}.

The main approximations of the  ISCT EoS  \cite{ISCT1,ISCT2} are  that the mean hard-core radius $\overline{R}= {\sum\limits_{l}\left\langle N_l\right\rangle R_l}\biggl/ {\sum\limits_{l}\left\langle N_l\right\rangle}$
 and the  mean  square of hard-core  radius $\overline{R^2} = \sum\limits_{k}  \langle N_k \rangle R_k^2 
 \biggl/ \sum\limits_{k} \langle N_k \rangle$ are self-consistently calculated via the average multiplicities $ \langle N_k \rangle$ of $k$-th sort of particles.  In this work we go one step further 
 and weaken these  assumptions. As it is shown below, now one can account for the fluctuations of the particle number density of each sort of particles.  Also 
 below it is shown that the alternative formulation of the ISCT EoS leads not to the exponential suppression of the surface tension and curvature tension coefficients, but to the power-like suppression which at low packing fractions  is similar to the density dependent interaction 
 recently introduced in Ref. \cite{Dutra1,Dutra2}.  Note that an accurate   accounting for the fluctuations of the particle number density and  charges  is extremely 
 important  nowadays in  view of the experimental  searches for   the (tri)critical endpoint of  the  phase diagram of quantum chromodynamics { \cite{Shuryak98}} which are going on at the  RHIC BNL and that are planned in the future experiments 
 at the   NICA JINR (Dubna) and at the  FAIR GSI (Darmstadt). 
 
 From  the  academic point of view it is also  rather  important that the method developed here provides more rigorous  derivation of the ISCT EoS which allows us to straightforwardly account for the third, the fourth and the fifth virial coefficients, compared to the heuristic derivation which was worked out in  \cite{ISCT1, ISCT2} (for more details see Sect. 2 below  { and Appendix A}).
 
 {
 The work is organized as follows. In the next Section we summarize the major approximations and 
 expressions of the standard ISCT EoS approach for the gas of  Boltzmann particles. 
 In Section 3 we compare  the alternative  formulation of the  ISCT EoS with the standard one
 and with the well-known  EoS of hard spheres for the one-component case. Such a comparison  allows us to 
further  generalize the alternative formulation of the  ISCT EoS both for the Boltzmann statistics and for the quantum ones.  Section 4 summarizes our conclusions. The Appendix  A  is devoted to the Laplace-Fourier transform method which allows us to weaken the assumptions of the standard ISCT EoS and to derive the alternative  formulation of the  ISCT EoS. 
 The expressions for the particle number density of the alternative formulation of the  ISCT EoS are given in the Appendix B.} 

\section{Self-consistent excluded volume approach  for multi-particle mixtures within the standard ISCT}

Consider the mixture of non-composite Boltzmann particles with the hard-core radii $\left\lbrace R_n ; n = 1, 2,..., N\right\rbrace $.
The composite particles whose constituents experience the hard-core repulsion with hadrons, light nuclei   were  considered recently in Refs. \cite{IST2b,IST2020a,IST2020b,IST2020c} and, hence, 
in the present work we do not pay much attention to their treatment.
For convenience, in this work the antiparticles are considered as the independent sorts. 
The excluded volumes per particle (classical second virial coefficients) $b_{kl}$
of particles $k$ and $l$ is given by
 \begin{equation}\label{Eq1}
	b_{kl} \equiv \frac{2}{3}\pi(R_k+R_l)^3 .
\end{equation}
Using it one can find the excluded volume of all pairs  taken per particle $\overline{V}_{excl}$   as
%
	\begin{equation}
	\label{Eq2}
	\overline{V}_{excl} = {\sum\limits_{k, l=1}^N N_k \frac{2}{3} \pi (R_k+R_l)^3 N_l}\biggl/ {\sum\limits_{k=1}^N N_k}  =  \frac{2}{3} \pi \frac{\sum\limits_{k, l=1}^N \hspace*{-1.56mm}N_k (R_k^3 +3R_k^2R_l+3R_kR_l^2+R_l^3) N_l}{\sum\limits_{k=1}^N N_k} ,
	\end{equation}
	where $N_k$ denotes the number of particles of sort $k$. The second equality in Eq. 
	(\ref{Eq2}) was obtained by opening the brackets 
 in the preceding expression.  Combining the 1-st  term with the 4-th one  in the numerator on the right-hand side of Eq. (\ref{Eq2}) and introducing the mean radius $\overline{R}$
 and the  mean square of hard-core  radius $\overline{R^2}$ as
\begin{eqnarray}\label{Eq3}
\overline{R} = {\sum\limits_{l=1}^N N_lR_l} \biggl/ {\sum\limits_{l=1}^N N_l} ,  \qquad 
\overline{R^2} \equiv 
	{\sum\limits_{k=1}^{N}N_k R_k^2}\biggl/{\sum\limits_{k=1}^{N}N_k}, 
\end{eqnarray}
one can identically rewrite the total excluded volume (\ref{Eq1}) 
in terms of eigenvolumes $V_k = \frac{4}{3} \pi R_k^3$, eigensurfaces $S_k = {4} \pi R_k^2$  and (double) eigenperimeter $C_k = {4} \pi R_k$ of particles of $k$-th sort 
as  \cite{ISCT1, ISCT2}
	\begin{eqnarray}\label{Eq4}	
\overline{V}_{excl} = \sum\limits_{k=1}^{N} N_k V_k +  A \cdot \overline{R}\sum\limits_{k=1}^{N} N_k S_k +
B \cdot \overline{R^2} \sum\limits_{k=1}^{N}N_k C_k ,
	\end{eqnarray}	
where the coefficients $A=0.5$ and $B=0.5$ are introduced for convenience.  Later on  we  will consider the coefficients $A$ and $B$ as the adjustable parameters. 

In order to evaluate the grand canonical partition function (GCPF) with  the  excluded volume   (\ref{Eq4}),  in our previous publications \cite{ISCT1, ISCT2}  it was  assumed that for large systems (in thermodynamic limit)  one can replace the quantities  $\overline{R}$ and  $\overline{R^2}$  by their values calculated  in  the  thermodynamic limit, i.e. 
\begin{eqnarray}\label{Eq5}
\overline{R} \underbrace{\longrightarrow}_{ V \rightarrow \infty} {\sum\limits_{l=1}^N\left\langle N_l\right\rangle R_l}\biggl/ {\sum\limits_{l=1}^N  \left\langle N_l\right\rangle} , \qquad 
	\overline{R^2}  \underbrace{\longrightarrow}_{ V \rightarrow \infty}  
	{\sum\limits_{k=1}^{N}\langle N_k \rangle R_k^2}\biggl/{\sum\limits_{k=1}^{N}\langle N_k \rangle},
\end{eqnarray}
where  the mean number of particles of $l$-th sort $\left\langle N_l\right\rangle$ is found  self-consistently after calculating  the GCPF under the assumptions given by Eqs. (\ref{Eq5}). 

Such an approach is very convenient to analyze the multi-component systems, since the resulting pressure $p$ and the coefficient of induced surface tension $\Sigma$ 
and the one of induced curvature tension $K$
form a closed system of three equations 
\begin{eqnarray}
\label{Eq6}
p &=&   T \sum\limits_{k = 1}^{N} \phi_k \exp \hspace*{-0.55mm}  \left[ \frac{\mu_k - V_k p - S_k \Sigma -C_k  K}{T}\right] ,
\\
\label{Eq7}
\Sigma &\equiv &  A \overline{R} \, p  =  A T \sum\limits_{k = 1}^{N} \hspace*{-0.55mm} R_k \phi_k \exp \hspace*{-0.55mm} \left[ \frac{\mu_k - V_k p - S_k \Sigma -C_k  K}{T}\right] ,
\\
\label{Eq8}
K &=&  B  \overline{R^2} \, p    = B T   \sum\limits_{k = 1}^{N}\hspace*{-0.55mm}  R_k^2 \phi_k \exp \hspace*{-0.55mm}  \left[ \frac{\mu_k - V_k p - S_k \Sigma -C_k  K}{T}\right]  , 
\end{eqnarray}
and this  number  of equations does not depend   on the number of different hard-core radii in the system. In contrast to ordinary formulation of the multi-component  VdW gas  EoS 
\cite{MVdW1,MVdW2,MVdW3} this property provides an essential numerical advantage 
\cite{IST1,IST2}.

Note that a similar equation for pressure  (\ref{Eq6}) was first postulated in Ref. \cite{Dillmann}
as a generalization of the famous Fisher droplet model \cite{Fisher67}. 
Then it was refined  in Ref.  \cite{Ford}, but, in contrast to our self-consistent treatment,  in Refs. 
\cite{Dillmann,Ford} and their followers the coefficients of surface $\Sigma$ and curvature  $K$ tensions were just the fitting parameters, while in our approach they are defined, respectively,  by Eqs. 
 (\ref{Eq7}) and  (\ref{Eq8}).

In Eqs. (\ref{Eq6})-(\ref{Eq8}) we introduced the chemical potential $\mu_k$  and  the thermal particle number density $\phi_k$ of the $k$-th sort of particles
in the system of volume $V$ which has the temperature $T$. 
Note that the thermal particle number density $\phi_k$
contains  the Breit-Wigner mass distribution. 
In the Boltzmann approximation  $\phi_k$  can be written as
\begin{eqnarray}\label{Eq9}
\phi_k = g_k  \gamma_S^{|s_k|} \int\limits_{M_k^{Th}}^\infty  \, 
\frac{ d m}{N_k (M_k^{Th})} 
\frac{\Gamma_k}{(m-m_{k})^{2}+\Gamma^{2}_{k}/4} \times  \int \frac{d^3 p}{ (2 \pi \hbar)^3 }   \exp \left[{\textstyle  - \frac{ \sqrt{p^2 + m^2} }{T} }\right] \,,
\end{eqnarray}
where $g_k$ is  the degeneracy factor of the $k$-th  sort of particle of  the mean mass 
in the  vacuum being $m_k$. In Eq. (\ref{Eq9})
$\gamma_S$ is the  strangeness suppression factor \cite{Rafelski} of these particles,  and 
$|s_k|$ is the number of valence strange quarks and antiquarks in this 
sort of particle.  The factor 
\begin{equation}\label{Eq10}
\displaystyle {N_k (M_k^{Th})} \equiv \int
\limits_{M_k^{Th}}^\infty \frac{d m \, \Gamma_k}{(m-m_{k})^{2}+
\Gamma^{2}_{k}/4} 
\end{equation}
denotes 
a normalization constant with    $M_k^{Th}$ being   the decay 
threshold mass of the $k$-th hadronic resonance, while  $\Gamma_k$ denotes its full   width  in the vacuum.
Evidently, for the stable hadrons  the  width $\Gamma_k$ should be set to zero,
which leads to the familiar expression for the thermal density
\begin{equation}\label{Eq11}
\phi_k \underbrace{\longrightarrow}_{\Gamma_k \rightarrow 0}  g_k \gamma_S^{|s_k|} \int  \frac{dp^3}{(2\pi\hbar)^3}  \exp \left[{\textstyle  - \frac{ \sqrt{p^2 + m^2_k} }{T} }\right] .
	\end{equation}
{ Although the Breit-Wigner ansatz for the resonance  mass distribution  is an approximation which
 is usually valid  for relatively 
narrow resonances only,  such an expression  for  thermal density  of  unstable particles (\ref{Eq9})   
provides a reasonable accuracy \cite{Hufner:1994ma,Wergieluk:2012gd,Blaschke:2013zaa,ResWidth,Kuksa}. 
Moreover,  for some dynamical models of hadron structure such as the Nambu--Jona-Lasinio model, one can  
split  the full spectral function of resonance  into the  resonant part  that 
corresponds to an unstable hadron state which   can be  approximated by a Breit-Wigner ansatz and the residual, repulsive part
\cite{Hufner:1994ma,Wergieluk:2012gd,Blaschke:2013zaa},  that   can be further 
approximated by the hard-core repulsion.  Therefore, we will employ  this ansatz for the quantum particles as well. 
}

 The VdW system (\ref{Eq6})-(\ref{Eq8})  can be refined further by making the following replacements  \cite{IST0, ISCT1, ISCT2}
	\begin{equation}
	\label{Eq12}
	\Sigma S_k \rightarrow \Sigma S_k \alpha_k, \quad {\rm with} \quad \alpha_k >1
	\,, \qquad K C_k \rightarrow \beta_k K C_k \quad {\rm with} \quad \beta_k >1 \,,
	\end{equation}
which lead to the standard   ISCT  EoS 
\begin{eqnarray}
\label{Eq13}
p &=&   \sum\limits_{k = 1}^{N} p_k  = T \sum\limits_{k = 1}^{N} \phi_k \exp \hspace*{-0.55mm}  \left[ \frac{\mu_k - V_k p - S_k \Sigma -C_k  K}{T}\right] ,
\\
\label{Eq14}
\Sigma &\equiv &    \sum\limits_{k = 1}^{N} \Sigma_k =  T A \sum\limits_{k = 1}^{N} \hspace*{-0.55mm} R_k \phi_k \exp \hspace*{-0.55mm} \left[ \frac{\mu_k - V_k p - S_k \alpha_k \Sigma -C_k  K}{T}\right] ,
\\
\label{Eq15}
K &=&     \sum\limits_{l = 1}^{N} K_l \hspace*{-0.55mm}=   T B  \sum\limits_{l = 1}^{N}\hspace*{-0.55mm}  R_l^2 \phi_l \exp \hspace*{-0.55mm}  \left[ \frac{\mu_k - V_l p - S_l \alpha_l \Sigma -C_l \beta_l K}{T}\right]  , 
\end{eqnarray}
where the particle pressure $p_k$, partial coefficient of surface tension $\Sigma_k$
and partial coefficient of curvature  tension $K_k$ of the $k$-th sort of particles  
are  introduced for convenience. 

The auxiliary parameters $\alpha_k$ and $\beta_k$ introduced  in Eqs. (\ref{Eq13})-(\ref{Eq15}) should be fixed  in such a way that they describe the higher virial coefficients.
In Ref. \cite{ISCT2} one can find several examples of the  two-component mixtures of classical  hard spheres and calssical  hard discs of different radii, which are very accurately described  by the system
(\ref{Eq13})-(\ref{Eq15})   practically in the whole gaseous phase, i.e. for the packing fraction 
$\eta =  \sum\limits_{k = 1}^{N} V_k  \rho_k  \le  0.45$  for hard spheres  (here $\rho_k$ denotes the particle number density of the $k$-th sort of particles)  and 
$\eta  \le  0.7$ for hard discs.

This framework has very clear physical grounds  to introduce the parameters $\alpha_k$ and $\beta_k$. From expression for the system pressure   (\ref{Eq13}) one can find the 
effective excluded volume of the $k$-th sort of particles as
\begin{eqnarray}\label{Eq16}
&& \hspace*{-5.5mm}V_k^{eff}  \equiv   \frac{\left[V_k p + S_k \Sigma + C_k K\right]}{p}   = 
V_k + A S_k \overline{R} + B C_k  \overline{R^2} = 
   \\
\label{Eq17}
& &\hspace*{-5.5mm} =  V_k + A S_k \frac{ \sum\limits_{l = 1}^{N}p_l R_l e^{-(\alpha_l-1)S_l \Sigma/T } }{ \sum\limits_{l = 1}^{N} p_l  } +  B C_k   \frac{ \sum\limits_{l = 1}^{N} p_l R_l^2 e^{ -(\alpha_l-1)S_l \Sigma/T  - (\beta_l-1)C_l K /T  } }{ \sum\limits_{l = 1}^{N}p_l } , ~
\end{eqnarray}
where the last result is obtained using Eqs.  (\ref{Eq14}) and (\ref{Eq15}) for $\Sigma_k$ and $K_k$, respectively. Note, however, that the average hard-core radius $\overline{R}$ and the average square of hard-core  radius   $\overline{R^2}$ in Eq.  (\ref{Eq16}) slightly differ from the corresponding values in Eqs. (\ref{Eq7}) and (\ref{Eq8}) due to the  assumptions (\ref{Eq12}). 

From Eq. (\ref{Eq16}) it is clear that $\widetilde  V_k^{eff}$  is the excluded volume, since it  stays in the exponential functions in Eqs. (\ref{Eq13})-(\ref{Eq15})  in front of the system pressure $p$.
 Eq. (\ref{Eq17}) leads to a gradual decrease of the effective excluded volumes $\{V_k^{eff}, k=1, 2, ... \}$,    if the system pressure increases.
Indeed, Eq.  (\ref{Eq17}) shows that for low densities, i.e. for $\Sigma S_k^{max}/T \ll 1$ and $K C_k^{max}/T \ll 1$, each exponential in Eq. (\ref{Eq17}) can be approximated as $\exp \left[- \frac{(\alpha_l-1)S_l \Sigma}{T} \right] \simeq 1$ and 
$\exp\left[ - \frac{(\beta_l-1)C_l K }{T}  \right] \simeq 1 $, which automatically  recover 
the usual multi-component VdW result for low densities \cite{ISCT1,ISCT2}. 
 However, for the limit of high pressure (or high densities) 
one can easily show  the validity of  the inequalities $\frac{\Sigma {S_k}}{T} \gg 1$ for  any $S_k >  0$  and  $\frac{K {C_k}}{T} \gg 1$ for  any $C_k >  0$. 
However,  under the conditions $\alpha_k > 1$ and  $\beta_k > 1$
one can see that the ratios $ \frac{\Sigma}{p}$ and  $ \frac{K}{p}$ vanish in this limit.
In other words, in the limit of high pressure the effective excluded 
volume of $k$-th sort of particles approaches their eigenvolume,  
i.e. $V_k^{eff} \rightarrow V_k  $. 

This is very convenient framework to use it for the GCPF, but one can make one step further,
namely to abandon the approximations (\ref{Eq5}) and consider the less strict approximations
\begin{eqnarray}\label{Eq18}
\overline{R}_{new}  =  \sum\limits_{l}  N_l  R_l \biggl/ \langle N \rangle  , \qquad 
	\overline{R^2}_{new}  = \sum\limits_{l}  N_l  R_l ^2\biggl/ \langle N \rangle  ,
\end{eqnarray}
where the quantity  $\langle N \rangle \equiv \sum\limits_{l}\left\langle N_l\right\rangle  \simeq \frac{p V}{T}$ can be found directly from the GCPF  (here $V$ is the volume of the system).   
This is a plausible approximation which is almost exact for low packing fractions $\eta < 0.1$,
but as it will be shown later  at higher packing fractions  the effects of suppression of the terms  that are proportional to 
$\overline{R}_{new}$ and $\overline{R^2}_{new}$
 will play more  important role and, hence, this approximative treatment will not be of a crucial importance. 
However,  such an approximation,  first of all, allows one to   greatly simplify the expression for $\langle N \rangle$ (see the corresponding expressions  for particle number densities in the Appendix) and,  second, as it will be shown later,  it  exactly corresponds to the results of Eqs. (\ref{Eq7}) and (\ref{Eq8}). Due to these  properties  it will be easier to demonstrate   all the additional findings  of the present approach  by  comparing  the new expressions with the   equations of the  standard ISCT EoS.

{
Apparently, the new approximations 
(\ref{Eq18}) allow one to more accurately account for the particle number fluctuations,
which is very important for the investigation of   event-by-event fluctuations in the context 
of the experimental searches for the (tri)critical endpoint of the strongly interacting matter phase diagram \cite{Shuryak98,Bugaev13,IST1b}. 
Moreover, as it will be shown below, the GCPF  evaluation  under the approximations 
(\ref{Eq18})  will lead to an alternative formulation of the ISCT EoS. }


In order to evaluate the GCPF of the considered mixture of Boltzmann particles
	\begin{eqnarray}\label{Eq19}
	& \hspace*{-1.1mm}Z_{GCE}(T,\left\lbrace \mu_k \right\rbrace, V) \equiv  \sum\limits_{ \left\lbrace { N_k} \right\rbrace }^\infty  \left[ \prod\limits_{k=1}^{N} \frac{\left[	\phi_k e^{\frac{\mu_k}{T} }(V - \overline{V}_{excl}) \right]^{N_k}}{N_k!} \right] \theta(V - \overline{V}_{excl}) , \\
	& \hspace*{-1.1mm} \overline{V}_{excl} =   \sum\limits_{k=1}^{N} N_k V_k +  A \cdot \overline{R}_{new}\sum\limits_{k=1}^{N} N_k S_k +
B \cdot \overline{R^2}_{new} \sum\limits_{k=1}^{N}N_k C_k .
	\label{Eq20}
	\end{eqnarray}
where $\phi_k$ is a thermal density of particles defined by Eqs. (\ref{Eq9}) and (\ref{Eq10}). 

Unfortunately, the direct evaluation of the system (\ref{Eq19}) and (\ref{Eq20})  is very difficult task, since the new mean hard-core radius $\overline{R}_{new}$
 and the  new mean square of hard-core  radius  $\overline{R^2}_{new}$ itself  depend on  all 
 particle occupational numbers  $\{N_k = 0, 1, 2, ..., k= 1, 2, ... \}$. In order to 
 evaluate the GCPF  (\ref{Eq19})  one has to get rid the $\Theta$-function constraint 
 which provides that the particle excluded volumes do not overlap and that the available volume for
 the motion of particles is nonnegative.   However, the usual Laplace transform technique 
 will not work in this case, since the excluded volume $\overline{V}_{excl} \sim \sum\limits_{k, l =1}^{N} N_k N_l $ in the GCPF  (\ref{Eq19}) is a nonlinear function of particle multiplicities $ N_k $.  
{ Fortunately, this difficulty can be overcome with the help of the Laplace-Fourier transform technique worked out  in Refs. \cite{LFtrans1, LFtrans2}. Since this evaluation is rather lengthy, we moved the technical details to Appendix A, while below we concentrate on the analysis of  analytical expression obtained  for the GCPF  (\ref{Eq19}).}

\section{Analysis of the alternative formulation of the ISCT EoS and its generalizations}\label{Sect3}

Introducing the statistical average of quantities  ${\cal O}_k$ as follows 
\begin{eqnarray}
   \hspace*{-6.3mm} \langle {\cal O} \rangle = \frac{ \sum\limits_{k=1}^N  {\cal O}_k\phi_k \exp \left[ \frac{\mu_k}{T}  - \frac{p}{T}(V_k+ AS_k\xi_1+BC_k\xi_2)  \right]  }{\sum\limits_{k=1}^N   \phi_k \exp \left[ \frac{\mu_k}{T}  - \frac{p}{T}(V_k+ AS_k\xi_1+BC_k\xi_2)  \right]  }  ,  
\end{eqnarray}
{one can  explicitly write  the  system of equations  derived  in  Appendix A from the GCPF  (\ref{Eq19})  in the  form}
	\begin{eqnarray}\label{Eq50}
	&& \hspace*{-6.3mm} 
	p = T \sum\limits_{k=1}^{N}  \phi_k \exp \left[ \frac{\mu_k}{T}  - \frac{p}{T}(V_k+ AS_k\xi_1+BC_k\xi_2)  \right]   ,  \quad {\rm where}\\
	&& \hspace*{-6.3mm} \xi_L = \frac{\Phi_L \left( \frac{p}{T} \right)}{p  \left[   1-    \frac{ \partial { \cal F   }(\lambda_0, \{ 0 \}) }{ \partial \lambda_0} \right]} \equiv 
  \frac{\langle {R^L} \rangle}{\left[   1 +  \frac{p \langle {V} \rangle}{T}  \left[ 1+ A \frac{\langle {S} \rangle}{\langle {V} \rangle} \xi_1 +   B \frac{\langle {C} \rangle}{\langle {V} \rangle} \xi_2 \right]  \right]}
	 , \quad {\rm for}\quad L=1, 2.
	 \label{Eq51}
	\end{eqnarray}
Comparing  Eqs. (\ref{Eq50}) and  (\ref{Eq51}) with the ISCT EoS  system (\ref{Eq13})-(\ref{Eq15}), one  can see that the expressions for system pressure are very similar. Moreover 
one can identify $\langle {R^L} \rangle$  in  Eqs. (\ref{Eq50}) and  (\ref{Eq51}) with   
$\overline{R^L}$ in  Eqs.  (\ref{Eq7})
 and (\ref{Eq8}), i.e. for the quantities derived within the VdW approximation.
However, in contrast to the standard ISCT EoS  system (\ref{Eq13})-(\ref{Eq15}),  the suppression of the 
surface  and curvature tensions due to the factors  $\xi_1$ and $\xi_2$ is not an exponential, but the power-like.

To solve the system  (\ref{Eq51}) for $\xi_1$ and $\xi_2$ we note that there is a useful relation
$\frac{\xi_2}{\xi_1} = \frac{\langle {R^2} \rangle}{\langle {R} \rangle} $, using which one can get 
an equation for $\xi_1$ in the convenient  form
	\begin{eqnarray}\label{Eq52}
	&& \hspace*{-6.3mm} \xi_1= 
  \frac{\langle {R} \rangle}{\left[   1 +  \frac{p \langle {V} \rangle}{T}  \left[ 1+ Q \xi_1 \right]  \right]}
	 , \quad {\rm where}\quad  Q \equiv A \frac{\langle {S} \rangle}{\langle {V} \rangle }  +    B \frac{\langle {C} \rangle \langle {R^2} \rangle}{\langle {V} \rangle \langle {R} \rangle} =   3(A+B)\frac{\langle {R^2} \rangle}{\langle {R^3} \rangle }  \, .
	\end{eqnarray}
Its physical solution  as a function of the dimensionless variable $z \equiv \frac{p \langle {V} \rangle}{T}$ is as follows 
	\begin{eqnarray}\label{Eq53}
	&& \hspace*{-6.3mm} \xi_1= 
  \frac{2 \langle {R} \rangle }{    1 +  z + \sqrt{(1+z)^2 + 4Qz \langle R \rangle   } } \quad \Rightarrow \quad  \xi_2= 
  \frac{2 \langle {R^2} \rangle }{    1 +  z + \sqrt{(1+z)^2 + 4Qz \langle R \rangle   } } \, , 
%
	\end{eqnarray}
although the relations for $\xi_1$ and $\xi_2$ contain the additional dependencies on $\xi_L$ through the statistical average values of the mean eigenvolume $\langle {V} \rangle $,  the mean eigensurface of particles $\langle {S} \rangle$ and  their mean eigenperimeter $\langle {C} \rangle $, Eqs. (\ref{Eq53}) are useful for qualitative analysis, since such additional dependencies are rather weak. 

{ Formally,  the right hand side of  the expressions  (\ref{Eq53}) can be interpreted in a way that  the  mean hard-core radius $\langle R \rangle$ and the mean hard core radius squared  $\langle R^2 \rangle$  are compressible.  This is in  a spirit of the works
\cite{Rozynek1,Rozynek2} in which the radius of nucleons in nuclear matter decreases, if  the pressure increases. Note, however, the principal difference between our approaches: in Refs. \cite{Rozynek1,Rozynek2} a single nucleon is considered microscopically as a kind of bag with the total radius of 0.7-0.8 fm, while we are studying the hard-core radii (for nucleons it is about 0.365 fm \cite{IST1,IST1b,IST1c,IST2}) for  an  ensemble of hadrons  which effectively decreases because at high  pressures  the particles  can be packed more densely,  than it is provided  by the usual VdW approximation.  As a result,  neither 
the eigenvolume $V_k$, nor the eigensurface $S_k$ and eigenperimeter $C_k$ of the particles of $k$-th sort are  modified, when the pressure increases.   
} 
	\begin{figure}[ht]
		\centerline{\includegraphics[scale=0.63]{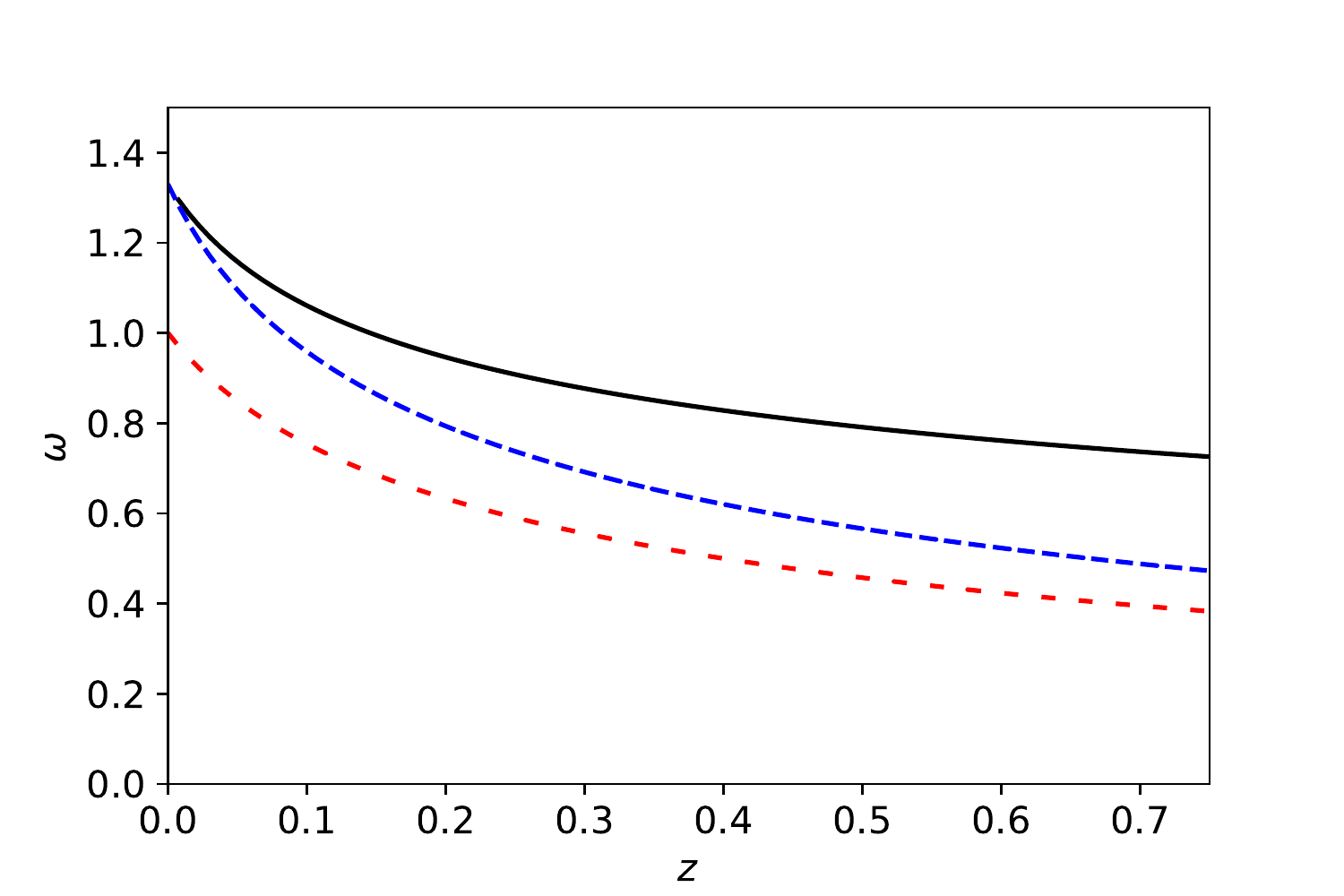}}
	\vspace*{-0.77mm}
	\caption{Comparison of  the effective excluded  volume suppression measures  $\omega_{St} $ (solid curve) and  $\omega_{Alt}$ (long dashed curve) for the set of parameters given by Eq. (\ref{Eq55}) (also shown in the legend).  For a comparison $\omega_{Alt}$ (short dashed curve) for
	$A=B=0.5$ is also shown. }
	\label{Fig1}
	\end{figure}


In order to compare the obtained alternative ISCT EoS (\ref{Eq50}) and  (\ref{Eq51}) with the standard one, we consider the one-component case, i.e. one sort of particles. In this case the statistical averaged of  the $n$-th power of hard-core radii  is  equal to the  $ R_1^n $ and  the quantity $Q$ from Eq. (\ref{Eq52}) acquires a simple form $Q= \frac{3(A+B)}{R_1}$.    Let us introduce the 
following measure of excluded volume suppression $\omega_{St}  $ for  the standard ISCT EoS  formulation and $\omega_{Alt}  $ for the alternative one 
\begin{eqnarray}\label{Eq54}
{
\omega_{St}   \equiv   \frac{S_1 \Sigma}{3 V_1 p} +  \frac{C_1 K}{3 V_1 p} }\qquad {\rm and} \qquad 
\omega_{Alt}   \equiv A \frac{S_1 \xi_1}{3 V_1} + B \frac{C_1 \xi_2}{3 V_1} =  \frac{(A+B)}{R_1}\xi_1 .
\end{eqnarray}
Evidently both of these measures show how the difference of the effective excluded  volume $V_1^{eff}$ (\ref{Eq16}) and the eigenvolume $V_1$ of particles depend on system pressure. 
In Fig. \ref{Fig1} one can see the comparison of  $\omega_{St} $ and  $\omega_{Alt}$ for the best set of parameters of  the system  (\ref{Eq13})-(\ref{Eq15})
\begin{eqnarray}\label{Eq55}
A = 0.57, \quad  B = 0.76,    \quad  \alpha = 1.07,  \quad  \beta = 3.76 ,
\end{eqnarray}
 which exactly reproduces  the five virial coefficients \cite{ISCT2}  of the famous CS   EoS \cite{CSEoS} of  classical hard spheres.  Note that the CS EoS   works  for hard spheres very well up to the packing fractions $\eta \simeq 0.45$ \cite{Simple_Liquids} { above which in the system of  classical hard spheres there exist  a phase transition to the solid phase.}

 {
As one can see from Fig. \ref{Fig1}  for  $z \le 0.2$ the maximal deviation between   $\omega_{Alt}$
and  $\omega_{St} $  is less than 20\%. Since for  $z \le 0.2$ the excluded volume effect is a correction to the ideal gas EoS, then the 20\% deviation of correction between them  is, indeed,  a tiny 
correction, which is hard to see.} 

To demonstrate this in a simple way  we consider the one-component system in the canonical ensemble, i.e. the pressure of the one-component system with the  suppression due $\xi_L$ factors from Eq.  (\ref{Eq53}) which  has the form
\begin{eqnarray}\label{Eq56}
p_{can}  \equiv  \frac{T \rho}{1 - \rho  \left[ V_1 + S_1 \xi \right]} =  \frac{T \rho}{1 - \rho V_1 \left[ 1 + \frac{3}{1+ \frac{V_1 p_{can} (A+B) }{T}}  \right]} ,
\end{eqnarray}
 where instead of the suppression factors  $\xi_1$ we used its approximation $\xi \equiv \frac{R_1}{1+ \frac{V_1 p_{can} (A+B) }{T}}$ which, apparently, works well not only at low and hight particle number density $\rho$. This reasonable approximation has an advantage that equation for the pressure 
$ p_{can}$ can be solved explicitly as
\begin{eqnarray}\label{Eq57}
p_{can}  \equiv  \frac{T }{2  V_1 (A+B)} \left[ \frac{(3+A+B) \eta}{1-\eta}-1 + \sqrt{ \left( 1-  \frac{(3+A+B) \eta}{1-\eta} \right)^2  +  4\frac{(A+B) \eta}{1-\eta}  }  \right]  ,
\end{eqnarray}
where the packing fraction is defined as $\eta = V_1 \rho$.
	\begin{figure}[th]
		\centerline{\includegraphics[scale=0.75]{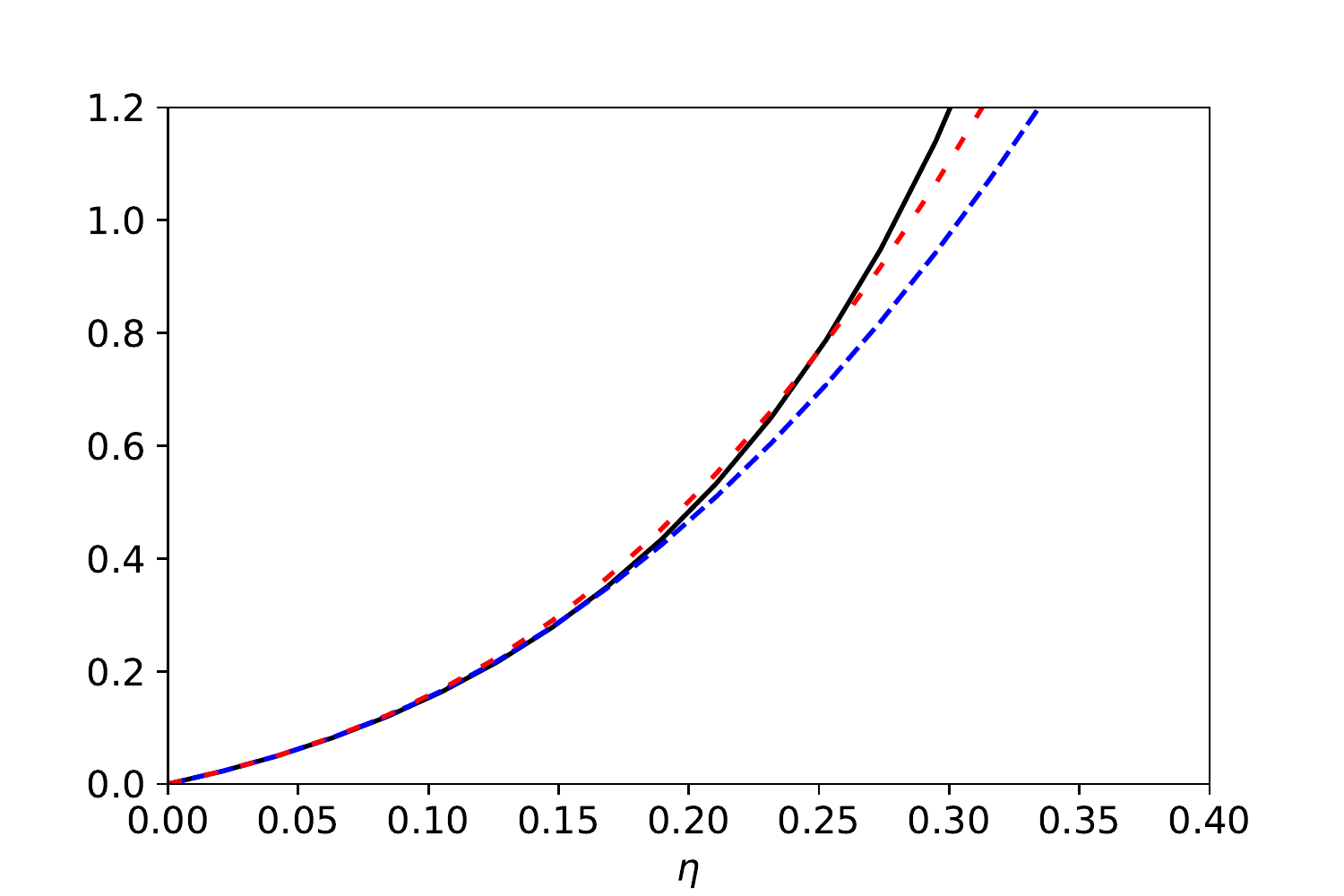}}
	\vspace*{-0.77mm}
	\caption{Comparison of  the dimensionless pressures  $\frac{p_{can}V_1}{T} $ for the CS EoS (solid curve) and  for the alternative ISCT EoS defined by Eq.  (\ref{Eq57})  for  the parameters $A=0.57$ and $B=0.76$ (long dashed curve) and  for   the parameters $A=0.5$ and $B=0.5$  (short dashed curve). }
	\label{Fig2}
	\end{figure}

	In Fig. \ref{Fig2} we compare the canonical ensemble pressure (\ref{Eq57}) (in a dimensionless form) with 
	{ the famous CS EoS  one  \cite{CSEoS} 
\begin{eqnarray}\label{Eq58}
\frac{p_{can}^{CS} V_1 }{T}  = \frac{\eta(1+\eta +\eta^2 - \eta^3) }{ (1 - \eta)^3}   ,
\end{eqnarray}
which nowadays is regarded as an exact result for the gaseous phase of  classical  hard spheres \cite{Simple_Liquids,Mulero}.}
As one can see from  Fig. \ref{Fig2}  the pressure (\ref{Eq57}) for the parameters $A=0.5$ and $B=0.5$ is less accurate for $\eta \in [0.1; 0.2]$, than  { the one  for the set with  $A=0.57$ and $B=0.76$, but the former  one can be used for  a wider range of packing fractions, namely for  $\eta \le 0.28$.  From  Fig. \ref{Fig2} one can see that  for packing fractions 
$\eta \le 0. 15$ the  canonical ensemble pressure (\ref{Eq57}) found  for the parameters $A=0.5$ and $B=0.5$ and 
the one found for  $A=0.57$ and $B=0.76$ are hardly distinguishable from each other and from the CS EoS.
}
	\begin{figure}[th]
		\centerline{\includegraphics[scale=0.75]{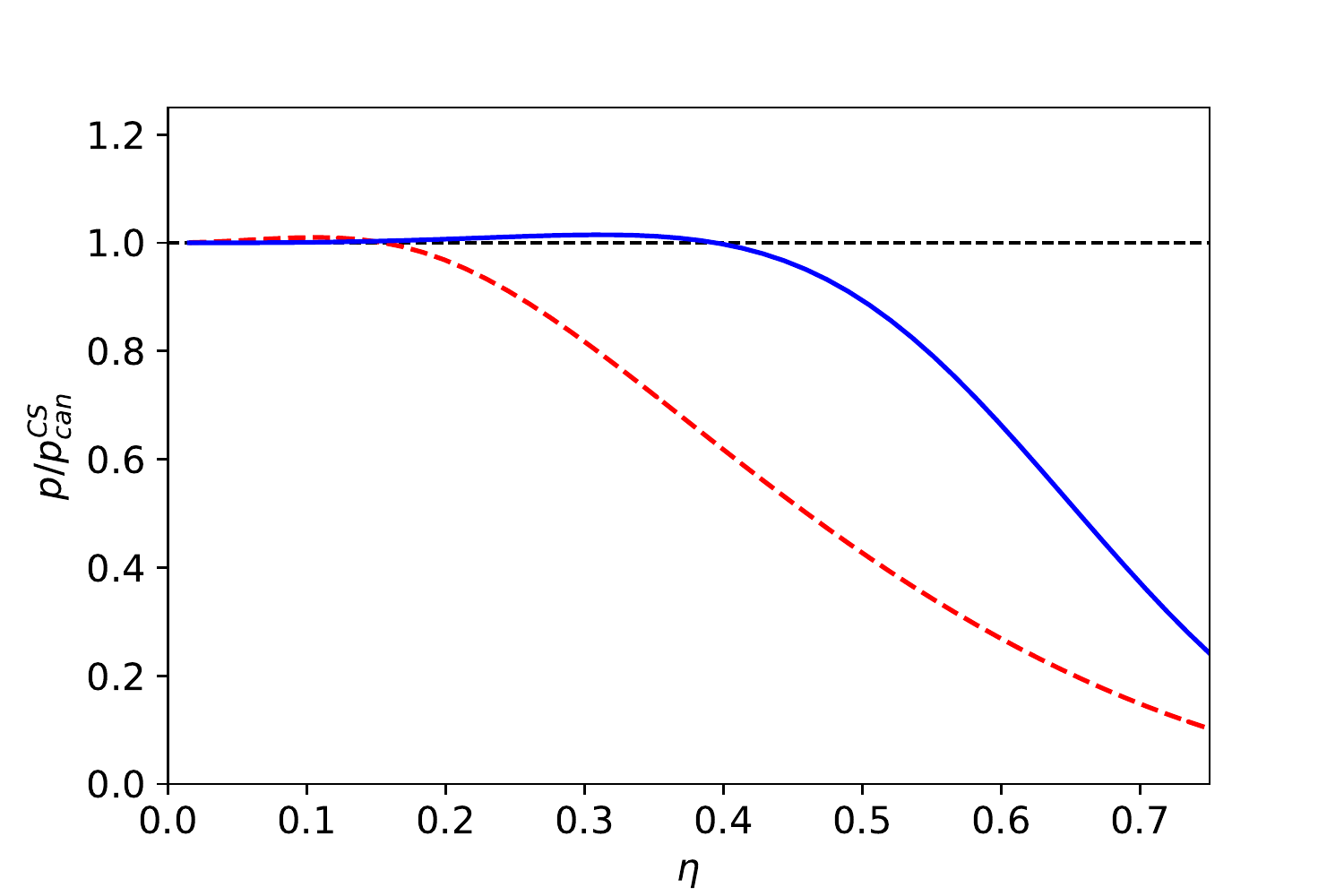}}
	\vspace*{-0.77mm}
	\caption{Packing fraction dependence of the ratios   $\frac{p_{can}}{p_{can}^{CS}}$ (long dashed curve is found for  the parameters $A=0.57$ and $B=0.76$) and  $\frac{p_{can}^G}{p_{can}^{CS}}$  (solid curve is calculated   for $\tilde A=0.0215$).  For $\eta \le 0.42$ the maximal relative deviation of  the ratio $\frac{p_{can}^G}{p_{can}^{CS}}$ from 1 does not exceed { 1.5\%.}  The short dashed curve $y=1$ is shown to guide an eye.  }
	\label{Fig3}
	\end{figure}

Here it is appropriate to { discuss}   the generalization of the canonical ensemble pressure  (\ref{Eq57})  in the spirit of the  Guggenheim EoS \cite{GuggenheimEOS} 
\begin{eqnarray}\label{Eq59}
\frac{p_{can}^G V_1}{T}  \equiv  \frac{ \eta}{\left(1 - \frac{ \eta }{4} \left[ 1 + \frac{3}{1+ \frac{p_{can}^G V_1   \tilde A }{T}}  \right] \right)^4}  \qquad {\rm with} \qquad  \tilde A  \simeq 0.0215 .
\end{eqnarray}
Evidently, for low packing fractions $\eta < 0.1$  the  suppression factor is about 1, but at higher packing fractions  the presence of suppression factor $\xi^G =  
 \frac{1}{1+ \frac{p_{can}^G V_1   \tilde A }{T} }$ in Eq.  (\ref{Eq59}) allows one to 
 extend  the validity range of the alternative ISCT EoS up to $\eta \le 0.42$ using just a single adjustable parameter $ \tilde A$.  Note that such a generalization automatically reproduces the value of the second and third virial coefficients and, hence, for this case the suppression should be  weaken.
   The  latter is seen from the small value of  the parameter $ \tilde A$.
  As one can see from   Fig. \ref{Fig3} the found generalization  (\ref{Eq59})   of the alternative ISCT EoS is able to 
accurately  reproduce the CS EoS with the maximal relative deviation not exceeding { 1.5\%} for 	$\eta \le 0.42$.  Hence such a generalization can be considered as an additional justification for the nonlinear 
suppression of interaction suggested recently in Ref. \cite{Dutra1,Dutra2}.

Now, however,  one can clearly  see the similarity  and difference of the present approach  with the density dependent interaction 
 suggested in Ref. \cite{Dutra1,Dutra2}.    At low densities,  both  the suppression factor $\xi$ and its 
generalization $\xi^G $ in Eq. (\ref{Eq59})  behave  as a power of packing fraction  $\eta$. However, 
at higher packing fractions one can replace the pressure $p_{can}$ and  $p_{can}^G$ by the CS one
to get an impression about the effect of excluded volume  suppression.  In the density dependent  interaction  approach   \cite{Dutra1} the maximal density is finite and, hence, at high packing fractions interaction does not disappear completely, while in the alternative ISCT EoS the pressure 
diverges in this limit and  the effect of interaction suppression can be stronger, especially for the 
generalized EoS  (\ref{Eq59}).  Moreover,  like the nonlinear suppression  of interaction suggested in  Ref.   \cite{Dutra1},  the generalized EoS (\ref{Eq59})  clearly demonstrates   that higher degree of suppression allows one to correctly reproduce the 
asymptotic behavior of hard-core  repulsion.

The success of the generalized EoS  (\ref{Eq59})  for the one-component system motivates  us  to 
generalize the alternative ISCT EoS (\ref{Eq50}), (\ref{Eq51}) in a similar  way 
	\begin{eqnarray}\label{Eq60}
	&& \hspace*{-6.3mm} 
	p^{new} = T \sum\limits_{k=1}^{N}  \phi_k \exp \left[ \frac{\mu_k}{T}  - \frac{p^{new}}{T}(V_k+ A_k S_k\xi_1+B_k C_k\xi_2)  \right]   ,  \quad {\rm where}\\
	&& \hspace*{-6.3mm} \xi_L  \equiv 
  \frac{\langle {R^L} \rangle}{\left[   1 + \gamma_1  \frac{p^{new} \langle {V} \rangle}{T}  \left[ 1+  \xi_1 \left( \frac{\langle {A\cdot S} \rangle}{\langle {V} \rangle}  +  \frac{\langle { B  \cdot C} \rangle \langle { S} \rangle }{\langle {V} \rangle \langle { C} \rangle } \right) \right]  \right]^{\gamma_2}}
	 , \quad {\rm for}\quad L=1, 2,
	 \label{Eq61}
	\end{eqnarray}
where in addition to the global  adjustable parameters $\gamma_1$ and  $\gamma_2$ we introduced 
the parameters $A_k$ and $B_k$ for the $k$-th sort of particles.   As it is argued in Refs. \cite{ISCT1,ISCT2} an introduction of the individual parameters $A_k$ and $B_k$ modifying 
the eigensurface and eigenperimeter  of  the standard ISCT EoS  may allow one to consider the  hard-core repulsion of 
non-spherical form { or  even to include the attraction between the constituents}.  Apparently, the same arguments  should be valid for the alternative 
ISCT EoS.

Another important application of the method developed here is an alternative ISCT EoS for quantum particles with the hard-core repulsion.  Applying the method presented in Appendix A  to the quantum GCPF of the model M2 from Ref.  \cite{ISCT1} one can straightforwardly obtain the quantum analog of alternative 
 ISCT  EoS  (\ref{Eq50}), (\ref{Eq51})  from  Eqs. (27) and (28) of Ref.  \cite{ISCT1}.  Then generalizing it in  the spirit of the system   (\ref{Eq60}), (\ref{Eq61}) one finds 
	\begin{eqnarray}\label{Eq62}
	&& \hspace*{-6.3mm} 
	p^q = T \sum\limits_{j=1}^{N}  g_j \hspace*{-1.4mm} \int\limits_{M_j^{Th}}^\infty  \hspace*{-1.4mm}
\frac{ d m}{N_j (M_j^{Th})} 
\frac{\Gamma_j}{(m-m_{j})^{2}+\Gamma^{2}_{j}/4} \times\hspace*{-1.4mm} \int \hspace*{-1.4mm} \frac{a_j  d^3 k }{(2 \pi \hbar)^3}   \ln \hspace*{-1.1mm}  \left[1 +  \frac{\exp\left[ \frac{\nu_j - e (k, m)}{T}  \right] }{ a_j}\right] ,   ~ {\rm with}  \qquad ~\\
	\label{Eq63}
	&& \hspace*{-6.3mm} 
	\nu_j \equiv \mu_j - \left[V_j +A_j S_j \xi_1 + B_j C_j \xi_2 \right]p^q \, ,  \quad {\rm and}\\
	&& \hspace*{-6.3mm} \xi_L  \equiv 
  \frac{\langle {R^L} \rangle}{\left[   1 + \gamma_1  \frac{p^q \langle {V} \rangle}{T}  \left[ 1+  \xi_1 \left( \frac{\langle {A\cdot S} \rangle}{\langle {V} \rangle}  +  \frac{\langle { B  \cdot C} \rangle \langle { S} \rangle }{\langle {V} \rangle \langle { C} \rangle } \right) \right]  \right]^{\gamma_2}}
	 , \quad {\rm for}\quad L=1, 2 \, .
	 \label{Eq64}
	\end{eqnarray}
Here the effective chemical  potential of $j$-th sort of particles is $\nu_j $, while $e (k, m)= \sqrt{m^2 + k^2}$ denotes the energy of such particles with momentum $\vec k$ and mass $m$ over which there is an integration with the Breit-Wigner weight.   Note that the expression for normalization factor $N_j (M_j^{Th})$ coincides with Eq. (\ref{Eq10}) and $M_j^{Th}$, as before,  is   the decay 
threshold mass of the $j$-th hadronic resonance which has the full  width $\Gamma_j$ in the vacuum.

The parameter $a_j =+1$  ($a_j =-1$) corresponds to the  Fermi-Dirac (Bose-Einstein) statistics.  One can readily check that in the limit $a_j \rightarrow \infty$ Eq. (\ref{Eq62})  automatically recovers the result (\ref{Eq60})  for the Boltzmann statistics.  The other notations are the same as for the  Boltzmann statistics case considered above. 

The coefficients $\gamma_1$, $\gamma_2$, $\{ A_k \}$ and $\{ B_k \}$ which are used in the systems   (\ref{Eq60}),  (\ref{Eq61}) and  (\ref{Eq62})-(\ref{Eq64}) can be found either from the comparison with the multi-component  version of the CS EoS, i.e. the MCSL EoS \cite{MCSL}, or from the analysis of  the molecular dynamics  results which can provide us with a few virial coefficients of multicomponent mixtures. 
After  these coefficients for the classical particles are found they can be used in the quantum version 
of alternative ISCT EoS. { The successful normalization of the standard ISCT EoS to various systems made in Ref. \cite{ISCT2} and the example of canonical pressure $p^G_{can}$ (\ref{Eq59}) compared above with the CS EoS  can be considered as  a perfect warranty that the
coefficients $\gamma_1$, $\gamma_2$, $\{ A_k \}$ and $\{ B_k \}$  of the alternative ISCT EoS  (\ref{Eq60}),  (\ref{Eq61})   can provide us with a very accurate description not only of the hard spheres with the multicomponent hard-core repulsion, but also of more complicated interaction between the particles. This, however, is not a trivial numerical task which require  a separate study.}

\section{Conclusions and perspectives}

In this work we developed a novel approach to analyze the excluded volume of the multicomponent mixtures of  classical hard spheres.  Using the Laplace-Fourier technique  \cite{LFtrans1, LFtrans2},  it was possible  to  evaluate the grand canonical partition function for the excluded volume which is quadratic in  the occupational number of particles under the plausible approximations.  To our best knowledge for the arbitrary number of different sorts of hard spheres 
this is done for the first time. 

As a result, the alternative formulation of the ISCT EoS has not an exponential, but the power-like suppression of the induced surface and curvature tension coefficients. Such formulation has some 
advantages over the standard  ISCT EoS formulation. In particular,  here it is  further  generalized  with the help of the  Guggenheim  trick used to obtain his famous EoS \cite{GuggenheimEOS}. 
For  the one-component case  it is shown that the generalized ISCT EoS with a single adjustable parameter  is able to reproduce 
the Carnahan-Starling EoS for the Boltzmann hard spheres with rather good accuracy up to packing fractions $\eta \simeq 0.42$. In the same way we obtain the  generalized ISCT EoS for the mixture of  quantum particles with the hard-core repulsion. 
 
It is necessary to stress that  the hard-core repulsion was considered here because it is  the simplest interaction which, nevertheless,  correctly  reflects the properties of  matter at high pressures.  It is clear that one can take into account  the attractive interaction using the same approach and obtain the power-like suppression. {Therefore, 
the present approach is rather general and, hence, it can be applied not only to develop the realistic EoS for the dense mixtures 
of  hadrons, nuclei and quark-gluon plasma bags  and the one  for  neutron  matter with a sizable fraction of  hyperons, but to 
resolve some old problems of dense quantum systems. For example, the classical hard spheres exhibit a deposition phase transition at packing fractions $\eta > 0.45$ \cite{Simple_Liquids,Mulero}, but to our best knowledge a similar problem
for quantum particles is not solved yet. Moreover,  recently it was shown  \cite{Bugaev:BEC20} that  Bose-Einstein condensation of non-relativistic  particles with the hard-core repulsion a la VdW  is a deposition phase transition similar to the system of classical hard-spheres
which belongs to the class of exactly solvable models of nuclear liquid-gas phase transitions  discusses in Ref. \cite{Bugaev+Reuter}. However, for an ultimate analysis of  this problem one needs a more realistic EoS which is valid at 
packing fractions $\eta > 0.45$.}

Moreover, one could  apply the Laplace-Fourier technique  to  the excluded volume of the multi-component mixtures in order to  completely abandon the approximative treatment of  
the mean hard-core radius and the mean square of  hard-core radius in Eqs. (\ref{Eq18}).
However, the resulting grand canonical partition function cannot be evaluated using the methods of the present work and, hence, this complicated mathematical task is reserved for the future exploration. 
It should be also mentioned that  after some modifications the present approach can be used not only in the thermodynamic limit, but for the finite volumes of the system, although  this task requires 
a separate analysis. 

\vspace*{2.2mm}

{\bf Acknowledgements.} { The author thanks Boris Grinyuk,  Oleksii Ivanytskyi, Pavlo Panasiuk, Oleksandr Vitiuk   and  Nazar Yakovenko for  fruitful discussions.  The author is thankful to  Nazar Yakovenko for verifying his calculations and for the help in improving the figures.
This work was supported  by the  National Academy of Sciences of Ukraine  under the Project No 0118U003197.} The author is  grateful to the COST Action CA15213 ``THOR`` for supporting his networking.

\setcounter{equation}{0}
\section{Appendix A:  Laplace-Fourier Transform Method}



\renewcommand{\theequation}{\rm{A}.\arabic{equation}}

{ In this Appendix we  evaluate the GCPF  (\ref{Eq19})   with the total excluded volume of particles  (\ref{Eq20})
using the Laplace-Fourier transform technique suggested in Refs. \cite{LFtrans1, LFtrans2}.}
 Our first step is to get rid of the nonlinear dependence of $\overline{V}_{excl}$  on $ N_k $ and  to equivalently make it linear in  $ N_k $.    For this purpose, prior to the Laplace transform of the GCPF (\ref{Eq19})  we will introduce in it the new variables $\xi_1$ and $\xi_2$ and use the following identity suggested in Refs. \cite{LFtrans1, LFtrans2}
 \begin{eqnarray}\label{Eq21}
G(x_1, x_2)  &\equiv&  \int\limits_{-\infty}^{+\infty} d \xi_1 \,  \delta (\xi_1- x_1) 
\int\limits_{-\infty}^{+\infty} d \xi_2 \,  \delta (\xi_2- x_2)  G(\xi_1, \xi_2) =  \\
\label{Eq22}
&= & \int\limits_{-\infty}^{+\infty} d \xi_1 \,   \int\limits_{-\infty}^{+\infty} \frac{d \eta_1}{2 \pi} \,   \int\limits_{-\infty}^{+\infty} d \xi_2 \,  \int\limits_{-\infty}^{+\infty} \frac{d \eta_2}{2 \pi}\,   e^{i\eta_1(\xi_1-x_1)} \,   e^{i\eta_2(\xi_2-x_2)}  \,  G(\xi_1, \xi_2)  \,,
\end{eqnarray}
where  Eq. (\ref{Eq22}) is obtained from  Eq. (\ref{Eq21}) with the help of the Fourier representation of the Dirac $\delta$-function. 

However, a word of caution should be said here. Since we are dealing with the products of 
generalized functions, it is necessary to define them very carefully before taking the thermodynamic limit $V$. Hence we will apply the identity  (\ref{Eq22}) in the meaning of 
principal value of each $\eta$-integral
 \begin{eqnarray}\label{Eq23}
G(x_1, x_2)  &= & \int\limits_{-\infty}^{+\infty} d \xi_1 \,   \int\limits_{-\infty}^{+\infty} d \xi_2 \, 
  \int\limits_{- {\cal A}_1}
^{+{\cal A}_1} \frac{d \eta_1}{2 \pi} \,  
  \int\limits_{-{\cal A}_2}^{+{\cal A}_2}  \frac{d \eta_2}{2 \pi}\,   e^{i\eta_1(\xi_1-x_1)} \,   e^{i\eta_2(\xi_2-x_2)}  \,  G(\xi_1, \xi_2)  \,.
\end{eqnarray}
Moreover,  after evaluation the GCPF  (\ref{Eq19}) for large, but finite values of parameters 
${\cal A}_1$ and ${\cal A}_2$, we will find the limits  ${\cal A}_1 \rightarrow +\infty$ and ${\cal A}_2  \rightarrow +\infty$ before taking the thermodynamic limit $V\rightarrow \infty$. 

Next  for the identity (\ref{Eq23}) we define the variables $x_1$ and $x_2$ as 
\begin{eqnarray}\label{Eq24}
x_1  =  \sum\limits_{l}  N_l  R_l \biggl/ \langle N \rangle  , \qquad 
	x_2  = \sum\limits_{l}  N_l  R_l^2 \biggl/ \langle N \rangle  .
\end{eqnarray}
Before making the Laplace transform of the GCPF (\ref{Eq19}) it is necessary to  note that 
the major task of this Appendix is to obtain an explicit expression for  such a partition as the function of two 
independent parameters, namely the
system volume $V$ and  the mean multiplicity of all particles $\langle N \rangle $.
Therefore, to reach this goal 
we   will consider the quantity  $\langle N \rangle $ as large, but the  volume independent parameter. 
The point is that we would like to exactly make all summations in  the system (\ref{Eq18})-(\ref{Eq20}) and get its exact representation in the thermodynamic limit while  keeping the quantity  $\langle N \rangle  $ as an independent parameter. 
Once the functional  dependence of the GCPF  (\ref{Eq19}) on the variables $V$ and  $\langle N \rangle  $
gets clear, we will easily find the thermodynamic limit using the volume dependence of $\langle N \rangle  $. Note, however, that the volume dependence of the quantity  $\langle N \rangle  $ can be, in principle,  taking into account  directly 
by adding the  $\xi_3$ and $\eta_3$ integrals in the identities  (\ref{Eq22}) and (\ref{Eq23}),
but this will produce unnecessary technical complications of the present work. Therefore, we 
will not introduce the additional variables $\xi_3$ and $\eta_3$ and consider  $\langle N \rangle $  as  a volume independent parameter till the analysis of the thermodynamic limit.

Under these assumptions with the help of the identity (\ref{Eq23}) we represent  the GCPF 
 (\ref{Eq19}) as 
	\begin{eqnarray}\label{Eq25}
	& \hspace*{-1.1mm}Z_{GCE}(T,\left\lbrace \mu_k \right\rbrace, V,  \langle N \rangle ) \equiv  
	\int\limits_{-\infty}^{+\infty} d \xi_1 \,   \int\limits_{-\infty}^{+\infty} d \xi_2 \, 
  \int\limits_{- {\cal A}_1}
^{+{\cal A}_1} \frac{d \eta_1}{2 \pi} \,  
  \int\limits_{-{\cal A}_2}^{+{\cal A}_2}  \frac{d \eta_2}{2 \pi}\,   \exp \left[  i\eta_1\xi_1 +  
  i\eta_2 \xi_2 \right]  \times  \,  \nonumber \\
& \hspace*{-1.1mm} \times 
\sum\limits_{ \left\lbrace { N_k} \right\rbrace }^\infty  \left[ \prod\limits_{k=1}^{N} \frac{\left( 	\phi_k \exp \left[ \frac{\mu_k}{T} - i \eta_1 \frac{R_k}{\langle N \rangle} -i \eta_2 \frac{R_k^2}{\langle N \rangle} \right]  \left[ V - \overline{V}_{excl}(\xi_1,\xi_2) \right]\right)^{N_k}}{N_k!} \right] \theta(V - \overline{V}_{excl}(\xi_1,\xi_2) ) , 
	\end{eqnarray}
where the total excluded volume now depends on the variables $\xi_1$  and $\xi_2$ as
	\begin{eqnarray}\label{Eq26}
	& \hspace*{-1.1mm} \overline{V}_{excl}(\xi_1,\xi_2) =   \sum\limits_{k=1}^{N} N_k V_k +  A \cdot \xi_1 \sum\limits_{k=1}^{N} N_k S_k +
B \cdot \xi_2 \sum\limits_{k=1}^{N}N_k C_k .
	\end{eqnarray}
Now we can make the Laplace transform of the GCPF  (\ref{Eq25}) with respect to system volume  while treating $\langle N \rangle  $ as a parameter.  The corresponding isobaric partition 
can be written as
	\begin{eqnarray}\label{Eq27}
	&&  \hspace*{-6.3mm} {\cal Z} (T,\left\lbrace \mu_k \right\rbrace, \lambda,  \langle N  \rangle ) \equiv  \int\limits_{0}^{+\infty} d V  \,  e^{- \lambda V } Z_{GCE}(T,\left\lbrace \mu_k \right\rbrace, V,  \langle N \rangle ) = \\
&& \hspace*{-6.3mm}= \int\limits_{0}^{+\infty} d V  \,  	\int\limits_{-\infty}^{+\infty} d \xi_1 \,   \int\limits_{-\infty}^{+\infty} d \xi_2 \, 
  \int\limits_{- {\cal A}_1}^{+{\cal A}_1} \frac{d \eta_1}{2 \pi} \,  
  \int\limits_{-{\cal A}_2}^{+{\cal A}_2}  \frac{d \eta_2}{2 \pi}\,   \exp \left[  i\eta_1\xi_1 +  
  i\eta_2 \xi_2  - \lambda V \right] \times   \,  \nonumber \\
&& \hspace*{-6.3mm} \times 
\sum\limits_{ \left\lbrace { N_k} \right\rbrace }^\infty  \left[ \prod\limits_{k=1}^{N} \frac{\left( 	\phi_k \exp \left[ \frac{\mu_k}{T} - i \eta_1 \frac{R_k}{\langle N \rangle} -i \eta_2 \frac{R_k^2}{\langle N \rangle} \right] \left[ V - \overline{V}_{excl}(\xi_1,\xi_2)  \right]\right)^{N_k}}{N_k!} \right] \theta(V - \overline{V}_{excl}(\xi_1,\xi_2) ) . \qquad 
\label{Eq28}
	\end{eqnarray}
Next we employ the usual trick (for an appropriate review see \cite{Bugaev+Reuter}) to evaluate
the summations in Eq.  (\ref{Eq28}) and to calculate the integral with respect to $V$, namely we change 
the order of integrals moving the $\int d \lambda$ to the right position and then we change 
the variable from $V$ to $\tilde V = V - \overline{V}_{excl}(\xi_1,\xi_2)$.  This change of variables
allows one to get rid of the $\Theta$-function constraint.  Moving the terms of $ \overline{V}_{excl}(\xi_1,\xi_2)$  containing the particle multiplicities $N_k$ to the appropriate places, after the described  change of variables one finds
	\begin{eqnarray}\label{Eq29}
	&&  \hspace*{-6.3mm} {\cal Z} (T,\left\lbrace \mu_k \right\rbrace, \lambda,  \langle N  \rangle )  =
	%
 	\int\limits_{-\infty}^{+\infty} d \xi_1 \,   \int\limits_{-\infty}^{+\infty} d \xi_2 \, 
  \int\limits_{- {\cal A}_1}^{+{\cal A}_1} \frac{d \eta_1}{2 \pi} \,  
  \int\limits_{-{\cal A}_2}^{+{\cal A}_2}  \frac{d \eta_2}{2 \pi}\, 
   \int\limits_{0}^{+\infty} d \tilde V   \, 
    \exp \left[  i\eta_1\xi_1 +  
  i\eta_2 \xi_2  - \lambda \tilde V \right] \times   \,  \nonumber \\
&& \hspace*{-6.3mm} \times 
\sum\limits_{ \left\lbrace { N_k} \right\rbrace }^\infty  \left[ \prod\limits_{k=1}^{N} \frac{\left( 	\phi_k \exp \left[ \frac{\mu_k}{T} - i \eta_1 \frac{R_k}{\langle N \rangle} -i \eta_2 \frac{R_k^2}{\langle N \rangle} - \lambda(V_k+ AS_k\xi_1+BC_k\xi_2)\right]  \tilde V \right)^{N_k}}{N_k!} \right] . \quad 
	\end{eqnarray}
Now  the summations over $N_k$ 
and product over $k$ can be made exactly, resulting in 
	\begin{eqnarray}\label{Eq30}
	&&  \hspace*{-6.3mm} {\cal Z} (T,\left\lbrace \mu_k \right\rbrace, \lambda,  \langle N  \rangle )  =
	%
 	\int\limits_{-\infty}^{+\infty} d \xi_1 \,   \int\limits_{-\infty}^{+\infty} d \xi_2 \, 
  \int\limits_{- {\cal A}_1}^{+{\cal A}_1} \frac{d \eta_1}{2 \pi} \,  
  \int\limits_{-{\cal A}_2}^{+{\cal A}_2}  \frac{d \eta_2}{2 \pi}\, 
   \int\limits_{0}^{+\infty} d \tilde V   \, 
    \exp \left[  i\eta_1\xi_1 +  
  i\eta_2 \xi_2 \right] \times   \,  \nonumber \\
&& \hspace*{-6.3mm} \times 
    \exp \left[ \tilde V   \sum\limits_{k=1}^{N}  \phi_k \exp \left[  f_k  (\lambda, \{ \eta_j \})  \right]  - \lambda \tilde V   \right]  ,  \quad  \\
	&&  \hspace*{-6.3mm}  {\rm where}~~f_k  (\lambda, \{ \eta_j \}) \equiv  \frac{\mu_k}{T} - i \eta_1 \frac{R_k}{\langle N \rangle} -i \eta_2 \frac{R_k^2}{\langle N \rangle} - \lambda(V_k+ AS_k\xi_1+BC_k\xi_2) . 
	\label{Eq31}
	\end{eqnarray}
Apparently, the integral with respect to the variable $\tilde V$ can be calculated easily and the isobaric partition (\ref{Eq27}) acquires the form
	\begin{eqnarray}\label{Eq32}
	&&  \hspace*{-6.3mm} {\cal Z} (T,\left\lbrace \mu_k \right\rbrace, \lambda,  \langle N  \rangle )  =
	%
 	\int\limits_{-\infty}^{+\infty} d \xi_1 \,   \int\limits_{-\infty}^{+\infty} d \xi_2 \, 
  \int\limits_{- {\cal A}_1}^{+{\cal A}_1} \frac{d \eta_1}{2 \pi} \,  
  \int\limits_{-{\cal A}_2}^{+{\cal A}_2}  \frac{d \eta_2}{2 \pi}\,    \, 
    \exp \left[  i\eta_1\xi_1 +  
  i\eta_2 \xi_2 \right]    \left[  \lambda -  \cal{F   (\lambda) }  \right]^{-1} ,  \qquad  \\
	&&  \hspace*{-6.3mm}  {\rm where}~~ { \cal{F   (\lambda, \{ \eta_j \} ) }}  \equiv   \sum\limits_{k=1}^{N}  \phi_k \exp \left[  f_k  (\lambda, \{ \eta_j \})  \right] . 
	\label{Eq33}
	\end{eqnarray}

Although all summations and the product  of   the original GCPF  (\ref{Eq25}) we calculated exactly 
the evaluation of the inverse Laplace transform of the isobaric  partition  (\ref{Eq32}) will be made under a plausible  approximation in order to simplify the presentation of the  major result of the present approach. 
 Since the functional $\lambda$-dependence
of the isobaric partition  (\ref{Eq32}) is similar to the standard case \cite{Bugaev+Reuter}, then
we conclude that in the thermodynamic limit the leading singularity (or the rightmost singularity) of 
the partition  (\ref{Eq32})  is the simple pole. The main difference, however, from the standard case is that this pole $\lambda_c = \lambda_0 + i \delta_1 \lambda$ is not the real one, but a  complex one. 

Now we recall the fact that the  variables $\eta_1 \in [- {\cal A}_1; + {\cal A}_1]$ and
 $\eta_2 \in [- {\cal A}_2; + {\cal A}_2]$ are restricted from above. Therefore, we can choose so large 
 values of $\langle N  \rangle$, i.e. very large volume of the system, that the quantities  
 \begin{eqnarray}\label{Eq34}
|\eta_1| \frac{R_k}{\langle N \rangle} \ll 1, \qquad |\eta_2| \frac{R_k^2}{\langle N \rangle} \ll 1,
\end{eqnarray}
can be made as small as necessary. Hence, one can make an expansion of the function ${ \cal F}   (\lambda_c)$ as follows
 \begin{eqnarray}\label{Eq35}
&&\hspace*{-4.4mm}{ \cal F   }(\lambda_c, \{ \eta_j \})   \simeq { \cal F  }  (\lambda_c, \{ 0\})  - \Phi_1 (\lambda_c)  i  \frac{\eta_1}{\langle N \rangle  }- \Phi_2 (\lambda_c)  i  \frac{\eta_2}{\langle N \rangle  } ,\\
&&\hspace*{-4.4mm}{\rm where}~~  \Phi_L (\lambda_c)  \equiv  T \sum\limits_{k=1}^N \frac{\partial { \cal F   }(\lambda_c, \{ 0 \}) }{\partial \mu_k}  R_k^L  = \sum\limits_{k=1}^N   R_k^L  \phi_k \exp \left[  f_k  (\lambda_c, \{ 0 \})  \right] , \quad {\rm for} \quad L=1, 2. \qquad 
\label{Eq36}
\end{eqnarray}
Using this expansion we can find the simple pole of the isobaric partition in the complex $\lambda$-plane.  

Due to inequalities (\ref{Eq34}) the  terms containing the ratios $\frac{\eta_1}{\langle N \rangle  }$ 
and $\frac{\eta_2}{\langle N \rangle  }$ which  are staying on the right hand side of Eq. (\ref{Eq35})
are small corrections to the term ${ \cal F  }  (\lambda_c, \{ 0\})$, hence the imaginary part of the pole $\lambda_c= \lambda_0 + i \delta_1 \lambda$, i.e. $\delta_1 \lambda$ is also small correction to $\lambda_0$. Therefore in solving the equation for $\lambda_c$
 \begin{eqnarray}\label{Eq37}
0= \lambda_c - { \cal F   }(\lambda_c, \{ \eta_j \})  \simeq \lambda_0 + i \delta_1 \lambda  - { \cal F   }(\lambda_c, \{ 0 \}) +  \Phi_1 (\lambda_c)  i  \frac{\eta_1}{\langle N \rangle  } + \Phi_2 (\lambda_c)  i  \frac{\eta_2}{\langle N \rangle  } ,
\end{eqnarray}
one can safely expand the function ${ \cal F   }(\lambda_c, \{ 0 \})$ in the vicinity of the point $\lambda=\lambda_0$, but at the same time one can also safely approximate $ \Phi_L (\lambda_c) \simeq \Phi_L (\lambda_0)$, since the term $\delta_1 \lambda$ will generate the higher order corrections which are not considered here (apparently, by a proper choice of $\langle N \rangle$ one can make the higher order corrections  negligible).   Taking this fact  into account one finds the equation for pole $\lambda_c$ as
 \begin{eqnarray}\label{Eq38}
&&   \hspace*{-14.5mm} 0= \lambda_c - { \cal F   }(\lambda_c, \{ \eta_j \})  \simeq  \nonumber \\
&&   \hspace*{-14.5mm} \simeq \lambda_0 + i \delta_1 \lambda  - 
{ \cal F   }(\lambda_0, \{ 0 \}) - \frac{ \partial { \cal F   }(\lambda_0, \{ 0 \}) }{ \partial \lambda_0} i  \delta_1 \lambda +  \Phi_1 (\lambda_0)  i  \frac{\eta_1}{\langle N \rangle  } + \Phi_2 (\lambda_0)  i  \frac{\eta_2}{\langle N \rangle  } ~ ~ \Rightarrow\\
&&  \hspace*{-14.5mm}   \lambda_0 - { \cal F   }(\lambda_0, \{ 0 \}) =0, ~~ {\rm and }~~
\delta_1 \lambda \simeq -  \left[ \Phi_1 (\lambda_0)    \frac{\eta_1}{\langle N \rangle  } + \Phi_2 (\lambda_0)   \frac{\eta_2}{\langle N \rangle  } \right]  \left[1 - \frac{ \partial { \cal F   }(\lambda_0, \{ 0 \}) }{ \partial \lambda_0} \right]^{-1}  . ~
   \label{Eq39}
\end{eqnarray}
These expressions above define the real part  of $\lambda_c$ as $\lambda_0 = Re(\lambda_c)$ and its imaginary part  as $\delta_1 \lambda = Im(\lambda_c)$. With their help one can easily the residue at  the pole $\lambda = \lambda_c$. Choosing  $\lambda = \lambda_R + i \delta_1 \lambda$ with $ \lambda_R \rightarrow  \lambda_0$, one can write  the right hand side Eq. (\ref{Eq37}) as
 \begin{eqnarray}\label{Eq40}
&& \hspace*{-11.2mm} \lambda_R + i \delta_1 \lambda  - { \cal F   }(\lambda_R + i \delta_1 \lambda, \{ 0 \}) +  \Phi_1 (\lambda_0)  i  \frac{\eta_1}{\langle N \rangle  } + \Phi_2 (\lambda_0)  i  \frac{\eta_2}{\langle N \rangle  } \rightarrow  \\
&& \hspace*{-11.2mm} \rightarrow   \lambda_R + i \delta_1 \lambda  - { \cal F   }(\lambda_0, \{ 0 \}) - 
  \frac{ \partial { \cal F   }(\lambda_0, \{ 0 \}) }{ \partial \lambda_0} [\lambda_R +i  \delta_1 \lambda
  - \lambda_0] 
   +  \Phi_1 (\lambda_0)  i  \frac{\eta_1}{\langle N \rangle  } + \Phi_2 (\lambda_0)  i  \frac{\eta_2}{\langle N \rangle  },   \qquad 
   \label{Eq41}
\end{eqnarray}
where in Eq (\ref{Eq40})  we expanded the function ${ \cal F   }(\lambda, \{ 0 \}) $ at the point 
$\lambda = \lambda_0$.  Applying the expressions (\ref{Eq39}) to Eq. (\ref{Eq41}), one finds the residue at the pole $\lambda_c$ as
 \begin{eqnarray}\label{Eq42}
Res\left( \left[ \lambda  - { \cal F   }(\lambda, \{ 0 \}) +  \Phi_1 (\lambda_0)  i  \frac{\eta_1}{\langle N \rangle  } + \Phi_2 (\lambda_0)  i  \frac{\eta_2}{\langle N \rangle  } \right]^{-1} \right)\Biggl|_{\lambda_c} =  \left[   1-    \frac{ \partial { \cal F   }(\lambda_0, \{ 0 \}) }{ \partial \lambda_0} \right]^{-1}.
\end{eqnarray}
With the help of last result  one can perform the inverse Laplace transform of the isobaric partition 
(\ref{Eq32}). Indeed,  choosing the integration contour in the complex $\lambda$-plane to be on the right hand side of $\lambda_c$, i.e.  with $\chi > \lambda_0$, one can 
make the inverse Laplace transform with respect to $\lambda$ in the following set of steps: first, 
we write  the definition 
	\begin{eqnarray}\label{Eq43}
	&&  \hspace*{-6.3mm} Z_{GCE}(T,\left\lbrace \mu_k \right\rbrace, V,  \langle N \rangle ) =
	 \int\limits_{\chi-i\infty}^{\chi+i\infty} \frac{d \lambda}{2 \pi i }  \,  e^{+ \lambda V } 
	{\cal Z} (T,\left\lbrace \mu_k \right\rbrace, \lambda,  \langle N  \rangle ) \simeq   \\
	&&  \hspace*{-6.3mm}	\simeq  
	 \int\limits_{\chi-i\infty}^{\chi+i\infty} \frac{d \lambda}{2 \pi i }  \,  e^{+ \lambda V } 
	 \int\limits_{-\infty}^{+\infty} d \xi_1 \,   \int\limits_{-\infty}^{+\infty} d \xi_2 \, 
  \int\limits_{- {\cal A}_1}^{+{\cal A}_1} \frac{d \eta_1}{2 \pi} \,  
  \int\limits_{-{\cal A}_2}^{+{\cal A}_2}  \frac{d \eta_2}{2 \pi}\, 
  {  \exp \left[  i\eta_1\xi_1 +  
  i\eta_2 \xi_2  + \lambda V \right] } \times \nonumber \\
	&&  \hspace*{-6.3mm}   \times \left[ \lambda  - { \cal F   }(\lambda, \{ 0 \}) +  \Phi_1 (\lambda_0)  i  \frac{\eta_1}{\langle N \rangle  } + \Phi_2 (\lambda_0)  i  \frac{\eta_2}{\langle N \rangle  } \right]^{-1} \, , 
\label{Eq44}
	\end{eqnarray}
next we change the order of integration and move the integral with respect to $\lambda$ to the 
rightmost position,  apply the Cauchy theorem on residues and get 
	\begin{eqnarray}\label{Eq45}
	 \hspace*{-1.5mm} Z_{GCE}(T,\left\lbrace \mu_k \right\rbrace, V,  \langle N \rangle ) 
	 && \hspace*{-5.5mm} \simeq 
	%
	\int\limits_{-\infty}^{+\infty} d \xi_1 \,   \int\limits_{-\infty}^{+\infty} d \xi_2 \, 
  \int\limits_{- {\cal A}_1}^{+{\cal A}_1} \frac{d \eta_1}{2 \pi} \,  
  \int\limits_{-{\cal A}_2}^{+{\cal A}_2}  \frac{d \eta_2}{2 \pi}\, 
  \frac{  \exp \left[  i\eta_1\xi_1 +  
  i\eta_2 \xi_2  + \lambda_c V \right] }{ \left[   1-    \frac{ \partial { \cal F   }(\lambda_0, \{ 0 \}) }{ \partial \lambda_0} \right]}  = \qquad  \\
&&\hspace*{-6.3mm} =	\int\limits_{-\infty}^{+\infty} d \xi_1 \,   \int\limits_{-\infty}^{+\infty} d \xi_2 \, 
  \int\limits_{- {\cal A}_1}^{+{\cal A}_1} \frac{d \eta_1}{2 \pi} \,  
  \int\limits_{-{\cal A}_2}^{+{\cal A}_2}  \frac{d \eta_2}{2 \pi}\, 
  \frac{  \exp \left[  i\eta_1\xi_1 +  
  i\eta_2 \xi_2  + \lambda_0 V  \right] }{ \left[   1-    \frac{ \partial { \cal F   }(\lambda_0, \{ 0 \}) }{ \partial \lambda_0} \right]}  \times \nonumber \\
  && \hspace*{-6.3mm}  \times  \exp \left[ - i V  \left[ \Phi_1 (\lambda_0)    \frac{\eta_1}{\langle N \rangle  } + \Phi_2 (\lambda_0)   \frac{\eta_2}{\langle N \rangle  } \right]  \left[1 - \frac{ \partial { \cal F   }(\lambda_0, \{ 0 \}) }{ \partial \lambda_0} \right]^{-1}  \right] ,~
  \label{Eq46}
	\end{eqnarray}
	where in the last step of evaluation we substituted the expressions (\ref{Eq39}) for $\lambda_c$ into Eq. (\ref{Eq45}).
	
Since $\lambda_0$ is related to the system pressure as $\lambda_0 = \frac{p}{T}$ (see Refs. \cite{Bugaev+Reuter,LFtrans1}), for large volume $V$ one can write   $\frac{V}{\langle N \rangle} = \frac{T}{p} = \lambda_0^{-1}$. Applying this relation  to  Eq.  (\ref{Eq46}),  one can now safely find  the limits ${\cal A}_1 \rightarrow + \infty$ and ${\cal A}_2 \rightarrow + \infty$ in  Eq.  (\ref{Eq46}), integrate over $\eta_1$ and $\eta_2$ and obtain 
	\begin{eqnarray}\label{Eq47}
	&& \hspace*{-6.3mm} Z_{GCE}(T,\left\lbrace \mu_k \right\rbrace, V,  \langle N \rangle ) 
	  \simeq  \int\limits_{-\infty}^{+\infty} d \xi_1 \,   \int\limits_{-\infty}^{+\infty} d \xi_2  \, 
	    \frac{  \exp \left[  \lambda_0 V  \right] }{ \left[   1-    \frac{ \partial { \cal F   }(\lambda_0, \{ 0 \}) }{ \partial \lambda_0} \right]} 
	   \times	\nonumber \\
&&\hspace*{-6.3mm}  \times	\,  
  \delta \left( \xi_1 - \frac{\Phi_1 (\lambda_0)}{\lambda_0 \left[   1-    \frac{ \partial { \cal F   }(\lambda_0, \{ 0 \}) }{ \partial \lambda_0} \right]} \right)  \delta \left( \xi_2 - \frac{ \Phi_2 (\lambda_0)}{\lambda_0  \left[   1-    \frac{ \partial { \cal F   }(\lambda_0, \{ 0 \}) }{ \partial \lambda_0} \right]} \right)  .
	\end{eqnarray}
Apparently, the integrations over $\xi_1$ and $\xi_2$ variables  are trivial  and, finally, the original GCPF (\ref{Eq25}) can be  now written  as 
	\begin{eqnarray}\label{Eq48}
&&\hspace*{-6.3mm} Z_{GCE}(T,\left\lbrace \mu_k \right\rbrace, V,  \langle N \rangle ) 
	  \simeq    \frac{  \exp \left[  \lambda_0 V  \right] }{ \left[   1-    \frac{ \partial { \cal F   }(\lambda_0, \{ 0 \}) }{ \partial \lambda_0} \right]} \Biggl|_{\xi_1 = \frac{\Phi_1 (\lambda_0)}{\lambda_0 \left[   1-    \frac{ \partial { \cal F   }(\lambda_0, \{ 0 \}) }{ \partial \lambda_0} \right]}; ~
\xi_2 = \frac{ \Phi_2 (\lambda_0)}{\lambda_0  \left[   1-    \frac{ \partial { \cal F   }(\lambda_0, \{ 0 \}) }{ \partial \lambda_0} \right]}} . 
	\end{eqnarray}
One can readily check that, if in Eq. (\ref{Eq48}) one assumes $R_k=0$ and $R_k^2 =0$ in functions $\Phi_L (\lambda_0)$  and ${ \cal F   }(\lambda_0, \{ 0 \})$ while keeping $V_k \neq 0$, the obtained result coincides with the usual GCPF of  particles with the   hard-core interaction in the limit
of eigenvolume approximation \cite{Bugaev+Reuter}. 

{ The final system of equations for  the pressure $p = T \lambda_0$ and for the quantities $\xi_1$ and $\xi_2$  
can be written  as
	\begin{eqnarray}\label{EqA29}
	&& \hspace*{-6.3mm} 
	p = T \sum\limits_{k=1}^{N}  \phi_k \exp \left[ \frac{\mu_k}{T}  - \frac{p}{T}(V_k+ AS_k\xi_1+BC_k\xi_2)  \right]   ,  \quad {\rm where}\\
	&& \hspace*{-6.3mm} \xi_L = \frac{\Phi_L \left( \frac{p}{T} \right)}{p  \left[   1-    \frac{ \partial { \cal F   }(\lambda_0, \{ 0 \}) }{ \partial \lambda_0} \right]} 	 , \quad {\rm for}\quad L=1, 2.
	 \label{EqA30}
	\end{eqnarray}
A detailed analysis of the derived EoS (\ref{EqA29}), (\ref{EqA30}) is made in  Section \ref{Sect3}.
}


\setcounter{equation}{0}

\section{Appendix B:  Expressions for the particle number density}

\renewcommand{\theequation}{\rm{B}.\arabic{equation}}

In this Appendix we summarize the expressions   for the particle number density of the systems  (\ref{Eq60}),  (\ref{Eq61}) and  (\ref{Eq62})-(\ref{Eq64}). First we consider the system (\ref{Eq60}),  (\ref{Eq61}) for the Boltzmann particles. Introducing the partial pressure 
$p_k^{new}$, the partial coefficients for the surface tension $\Sigma_k^{new}$ and the partial coefficients for the curvature tension $K_k^{new}$ of the $k$-th sort of particles, one can rewrite the  system (\ref{Eq60}),  (\ref{Eq61}) as
	\begin{eqnarray}\label{EqB1}
	&& \hspace*{-6.3mm} 
	p^{new} \equiv  \sum\limits_{k=1}^{N} p_k^{new} = T \sum\limits_{k=1}^{N}  \phi_k \exp \left[ \frac{\mu_k - V_k p^{new} + A_k S_k \Sigma^{new} +B_k C_k K^{new}}{T}  \right]   ,  \quad \\
\label{EqB2}
	&& \hspace*{-6.3mm} \Sigma^{new} \equiv  
	\sum\limits_{k=1}^{N} \Sigma_k^{new} = \frac{1}{G^{\gamma_2}} \sum\limits_{k=1}^{N} R_k p_k^{new}
,  \quad \\
\label{EqB3}
	&& \hspace*{-6.3mm} K^{new} \equiv  \sum\limits_{k=1}^{N} K_k^{new} = \frac{1}{G^{\gamma_2}} \sum\limits_{k=1}^{N} R_k^2  p_k^{new}
  ,  \quad \\
	 && \hspace*{-6.3mm} G   \equiv  1+ \gamma_1 \sum\limits_{j=1}^{N} F_j
	 \equiv 1+ \gamma_1 \sum\limits_{j=1}^{N}  \frac{( V_j p^{new} + A_j S_j \Sigma^{new} +B_j C_j K^{new})}{T} .
\label{EqB4}
	\end{eqnarray}
Differentiating Eqs. (\ref{EqB1})-(\ref{EqB3}) of this Appendix  with respect to the chemical potential $\mu_k$, one
can find the particle number density of the $k$-th sort of particles 
$\rho_k =  \frac{\partial p^{new} }{\partial \mu_k} $ and the derivatives  
$ \frac{\partial \Sigma^{new} }{\partial \mu_k}$  and  
$ \frac{\partial K^{new} }{\partial \mu_k}$. Solving the system of three equations for these quantities, one can find the particle number density of the $k$-th sort of particles 
$\rho_k$ as
\begin{equation}\label{EqB5}
\hspace*{-2.mm}\rho_k = \frac{1}{T} \frac{(e j -  f h ) B^{(1)}_k - (b j  - c h)B^{(2)}_k + (b f - c e) B^{(3)}_k}{a (e j -  f h ) - d (b j  - c h)  + g (b f - c e) } \,,
\end{equation}
where the coefficients $a, b, c, d, e, f, g, h$, $j$, $B^{(1)}_k$, $B^{(2)}_k$ and $B^{(3)}_k$ are 
given by
\begin{eqnarray}\label{EqB6}
&&a= 1 + \sum_k  V_k\frac{p_k^{new}}{T} , \quad \quad b = \sum_k  A_k S_k \frac{p_k^{new}}{T} ,  \qquad ~   c = \sum_k  B_k C_k \frac{p_k^{new} }{T}  , \\
\label{EqB7}
&&d = \sum_k  V_k \frac{B^{(2)}_k}{T}   ,  
\qquad \quad~ e =1 + \sum_k  A_k S_k \frac{B^{(2)}_k }{T}   , \quad f = \sum_k  B_k C_k  \frac{ B^{(2)}_k }{T} ,\quad  \\
\label{EqB8}
&& g = \sum_k  V_k \frac{B^{(3)}_k}{T}   ,  
\qquad \quad~ h = \sum_k  A_k S_k \frac{ B^{(3)}_k}{T}   , \qquad ~ j = 1+\sum_k 
B_k C_k   \frac{B^{(3)}_k}{T}  , \\
 \label{EqB9}
&&B^{(1)}_k  = p_k^{new},   \qquad B^{(2)}_k  = \Sigma_k^{new} - \frac{\gamma_1\gamma_2 \Sigma^{new}}{G} F_k , \qquad  B^{(3)}_k =  K_k^{new} - \frac{\gamma_1\gamma_2 K^{new}}{G} F_k . ~
%
\end{eqnarray}

Writing the quantum ISCT EoS  (\ref{Eq60})-(\ref{Eq62}) similarly to the system  (\ref{EqB1})-(\ref{EqB4}) of this Appendix
	\begin{eqnarray}\label{EqB10}
	&& \hspace*{-6.3mm} 
	p^{q} \equiv T \sum\limits_{j=1}^{N}  g_j  \hspace*{-1.4mm} \int\limits_{M_j^{Th}}^\infty  \hspace*{-1.4mm}
\frac{ d m}{N_j (M_j^{Th})} 
\frac{\Gamma_j}{(m-m_{j})^{2}+\Gamma^{2}_{j}/4}  \hspace*{-1.4mm} \int \hspace*{-1.4mm} \frac{a_j  d^3 k }{(2 \pi \hbar)^3}   \ln \hspace*{-1.1mm}  \left[1 +  \frac{1}{ a_j} \exp\left[ \frac{\nu_j - e(k, m)}{T}  \right] \right] ,    \quad  \\
\label{EqB11}
	&& \hspace*{-6.3mm} \Sigma^{q} \equiv  
	\sum\limits_{k=1}^{N} \Sigma_k^{q} = \frac{1}{(G_q)^{\gamma_2}} \sum\limits_{k=1}^{N} R_k p_k^{q}
,  \quad \\
\label{EqB12}
	&& \hspace*{-6.3mm} K^{q} \equiv  \sum\limits_{k=1}^{N} K_k^{q} = \frac{1}{(G_q)^{\gamma_2}} \sum\limits_{k=1}^{N} R_k^2  p_k^{q}
  ,  \quad \\
\label{EqB13}
	 && \hspace*{-6.3mm} G_q   
	 \equiv 1+ \gamma_1 \sum\limits_{j=1}^{N}  \frac{( V_j p^{q} + A_j S_j \Sigma^{q} +B_j C_j K^{q})}{T} , \\
\label{EqB14}
	 && \hspace*{-6.3mm} 	\nu_k \equiv \mu_k - \left[V_k p^q  +A_k S_k \Sigma^q + B_k C_k K^q \right]\, , 
	\end{eqnarray}
one can easily show that  Eq. (\ref{EqB5}) remains valid for the quantum case as well, if one makes 
the necessary replacements in Eqs.  (\ref{EqB6})-(\ref{EqB9})  
\begin{eqnarray}\label{EqB15}
&&\rho_k^q = \frac{1}{T} \frac{(e j -  f h ) B^{(q1)}_k - (b j  - c h)B^{(q2)}_k + (b f - c e) B^{(q3)}_k}{a (e j -  f h ) - d (b j  - c h)  + g (b f - c e) } \,,\\
\label{EqB16}
&&a= 1 + \sum_k  V_k n_k^{q} , \quad \quad b = \sum_k  A_k S_k n_k^{q},  \qquad ~   c = \sum_k  B_k C_k n_k^{q}   , \\
\label{EqB17}
&&d = \sum_k  V_k \frac{B^{(q2)}_k}{T}   ,  
\qquad \quad~ e =1 + \sum_k  A_k S_k \frac{B^{(q2)}_k }{T}   , \quad f = \sum_k  B_k C_k  \frac{ B^{(q2)}_k }{T} , \quad  \\
\label{EqB18}
&& g = \sum_k  V_k \frac{B^{(q3)}_k}{T}   ,  
\qquad \quad~ h = \sum_k  A_k S_k \frac{ B^{(q3)}_k}{T}   , \qquad ~ j = 1+\sum_k 
B_k C_k   \frac{B^{(q3)}_k}{T} ,   \\
 \label{EqB19}
&&B^{(q1)}_k  = T n_k^{q},   ~\quad B^{(2)}_k  = T\left[
 \frac{R_k n_k^{q}}{G_q^{\gamma_2}} - \frac{\gamma_1\gamma_2 \Sigma^q}{G_q} H_k^q  \right], \quad  B^{(3)}_k = T\left[   \frac{R_k^2 n_k^{q}}{G_q^{\gamma_2}}  - \frac{\gamma_1\gamma_2 K^q}{G_q} H_k^q  \right] , \qquad  \\
 \label{EqB20}
&&  H_k^q  \equiv ( V_k   +A_k S_k R_k + B_k C_k R_k^2 )n_k^q = V_k  (1+1.5 A_k + 1.5 B_k) n_k^q  ,
\end{eqnarray}
where $n_k^q$ denotes  the particle number  density of quantum particles of the $k$-th sort  with the effective chemical potential  $\nu_k$. It can be written explicitly as 
\begin{eqnarray} \label{EqB21}
 n_j^q  & \equiv & \frac{\partial p_j^q (T, \nu_j)}{\partial \nu_j } =  \nonumber \\
& = & g_j \hspace*{-1.4mm} \int\limits_{M_j^{Th}}^\infty  \hspace*{-1.4mm}
\frac{ d m}{N_j (M_j^{Th})} 
\frac{\Gamma_j}{(m-m_{j})^{2}+\Gamma^{2}_{j}/4} \times \hspace*{-1.4mm} \int \hspace*{-1.4mm} \frac{  d^3 k }{(2 \pi \hbar)^3}    \left[\exp\left[ \frac{ e (k, m) - \nu_j }{T}  \right]  +  \frac{1}{ a_j} \right]^{-1} . \quad 
 \end{eqnarray}


\vspace*{2.2mm}



\centerline{\bf \large References}



\begin{thebibliography}{99}

\bibitem{Bugaev07}
	%
	K. A. Bugaev, 
Phys. Rev. {\bf C 76},     014903 (2007);  
arXiv:0703222 [hep-ph] and references therein.

\bibitem{Bugaev+Reuter} 
	%
K. A. Bugaev and P. T. Reuter,
 {Ukr. J. Phys.} {\bf 52}, 489 (2007);
[arXiv:1001.4477 [nucl-th]].

\bibitem{BugaevFWM} 
	%
K. A. Bugaev, V. K. Petrov and G. M. Zinovjev,
{Phys. Rev.}  {\bf C  79},   054913--1-12  (2009).
 
 \bibitem{Alexei12a}
 %
A. I. Ivanytskyi and K. A. Bugaev,
{ Ukr. J. Phys.} {\bf 57},  964 (2012) and references therein.

\bibitem{Alexei12b}
%
A. I. Ivanytskyi, K. A. Bugaev, A. S. Sorin and G. M. Zinovjev,
Phys. Rev. E {\bf 86},  061107 (2012)
and references therein.
 
 
\bibitem{Bugaev13} 
%
K. A. Bugaev, V. K. Petrov and G. M. Zinovjev,
Phys.  Atom. Nucl.  {\bf 76},  341  (2013)  and references therein.


\bibitem{Simple_Liquids}
%
J. P. Hansen and I. R. McDonald, {\it  Theory of Simple Fluids} (Academic Press, Amsterdam, 2006).

\bibitem{Mulero}
%
{\it Theory and Simulation of Hard Sphere Fluids and Related Systems}, Lect. Notes Phys. Vol. 753, edited by A.
Mulero (Springer-Verlag, Berlin, 2008).

\bibitem{IST0}
%
V. V. Sagun, A. I. Ivanytskyi, K. A. Bugaev and I. N.
Mishustin, Nucl. Phys. A {\bf 924}, 24 (2014).

\bibitem{IST1}
%
K. A. Bugaev  et al., 
Nucl. Phys. A {\bf 970}, 133 (2018).

\bibitem{IST1b}
%
  K. A. Bugaev et al., 
 Phys. Part. Nucl. Lett. {\bf 15},  210-224 (2018).
 
 \bibitem{IST1c}
%
 A.~I.~Ivanytskyi, K.~A.~Bugaev, V.~V.~Sagun, L.~V.~Bravina and E.~E.~Zabrodin,
 Phys. Rev.  C {\bf 97},  064905--1-8 (2018).



\bibitem{IST2}
%
V. V. Sagun et al., 
 Eur. Phys. J. A {\bf 54},  100 (2018).

\bibitem{IST2a}
%
K. A. Bugaev, A. I. Ivanytskyi, V. V. Sagun, E. G. Nikonov and G. M. Zinovjev,
Ukr. J. Phys.  {\bf 63},  863-880 (2018). 

{
\bibitem{Veta19}
%
V.~V.~Sagun, I.~Lopes and A.~I.~Ivanytskyi,
Astrophys. J. \textbf{871},   157 (2019). 

\bibitem{Veta20}
%
V.~Sagun, G.~Panotopoulos and I.~Lopes,
Phys. Rev. D \textbf{101},   063025 (2020) and references therein.
}

\bibitem{IST2b}
%
K. A. Bugaev et al., 
J. of Phys. Conf. Series {\bf  1390},  012038, p. 1-6  (2019).

\bibitem{IST2020a}
%
B. E. Grinyuk et al., 
arXiv:2004.05481v1 [hep-ph] (2020) p. 1-15. { (to appear in Int. J. Mod. Phys. A)}

\bibitem{IST2020b}
%
K. A. Bugaev et al., 
Eur. Phys. J. A 56, 293--1-15
(2020) (doi.org/10.1140/epja/s10050-020-00296-5);
arXiv:2005.01555v1 [nucl-th].


\bibitem{IST2020c}
%
O.~V.~Vitiuk et al., 
arXiv:2007.07376 [hep-ph] (2020) p. 1-12.

\bibitem{ProtonFlow}
%
P. Danielewicz, R. Lacey, and W. G. Lynch, Science 298, 1593
(2002).

\bibitem{HRGM1}
 M.  Albright, J. Kapusta  and C. Young,  Phys. Rev. C {\bf 90}, 024915 (2014).
 
 \bibitem{HRGM2}
 S. Chatterjee et al., Adv. High Energy Phys. {\bf 2015}, 349013
(2015).
 
\bibitem{HRGM3} 
 A. Andronic, P. Braun-Munzinger, K. Redlich and J.
Stachel, J. Phys. Conf. Ser. {\bf 779}, 012012 (2017).
 

\bibitem{ISCT1}
%
K. A. Bugaev, 
 Eur. Phys. J. A {\bf 55}, 215  (2019).


\bibitem{ISCT2}
%
N. S. Yakovenko, K. A. Bugaev, L.V. Bravina and E. E. Zabrodin,
  arXiv:1910.04889 [nucl-th] (2019)  p. 1-13  { (to appear in EPJ ST)}

	\bibitem{CSEoS}
	%
	N. F. Carnahan and K. E. Starling, J. Chem. Phys. \textbf{51}, 635 (1969).
	
	\bibitem{MCSL}
	%
	G. A. Mansoori, N. F. Carnahan, K. E. Starling, T. Leland, J. Chem. Phys. \textbf{54}, 1523, (1971).

\bibitem{Dutra1}
%
O.~Louren\c{c}o, M.~Dutra, C.~H.~Lenzi, M.~Bhuyan, S.~K.~Biswal and B.~M.~Santos,
arXiv:1908.05114 [nucl-th] (2019) p. 1-11.

{
\bibitem{Dutra2}
M.~Dutra, B.~M.~Santos and O.~Louren\c{c}o,
J. Phys. G \textbf{47},   035101 (2020).

\bibitem{Shuryak98}
%
M. A. Stephanov, K. Rajagopal and E. V. Shuryak, 
Phys. Rev. Lett. {\bf 81}, 4816 (1998). 
}

\bibitem{MVdW1}
%
G. Zeeb, K. A. Bugaev, P. T. Reuter and H. St\"ocker,
{Ukr.  J. Phys.}  {\bf 53},      279-295 (2008).

\bibitem{MVdW2}
%
D. R. Oliinychenko, K. A. Bugaev and A. S. Sorin, 
Ukr. J. Phys.  {\bf 58},  211-227  (2013).

\bibitem{MVdW3}
%
K. A. Bugaev, D. R. Oliinychenko,   A. S. Sorin and G. M. Zinovjev, 
Eur. Phys. J. A {\bf 49}, 30  (2013). 

\bibitem{Dillmann}
%
A. Dillmann, G. E. Meier, J. Chem. Phys. {\bf 94}, 3872 (1991).


\bibitem{Fisher67}
%
M. E. Fisher, Physics {\bf 3},  255 (1967).

\bibitem{Ford}
%
A. Laaksonen, I. J. Ford, M. Kulmala, Phys. Rev. E {\bf 49}, 5517 (1994).

\bibitem{Rafelski}
J. Rafelski, Phys. Lett. B {\bf 62}, 333 (1991).

\bibitem{Hufner:1994ma} 
  J.~H\"ufner, S.~P.~Klevansky, P.~Zhuang and H.~Voss,
  Annals Phys.\  {\bf 234}, 225 (1994).


\bibitem{Wergieluk:2012gd} 
  A.~Wergieluk, D.~Blaschke, Y.~L.~Kalinovsky and A.~Friesen,
  Phys.\ Part.\ Nucl.\ Lett.\  {\bf 10}, 660 (2013).

\bibitem{Blaschke:2013zaa} 
  D.~Blaschke, M.~Buballa, A.~Dubinin, G.~R\"opke and D.~Zablocki,
  Annals Phys.\  {\bf 348}, 228 (2014) and references therein.


\bibitem{ResWidth} 
%
K. A. Bugaev,  A. I. Ivanytskyi,  D. R. Oliinychenko, E. G. Nikonov, V. V. Sagun   and G. M. Zinovjev,
Ukr. J. Phys. {\bf 60},  181  (2015) and  references therein.

\bibitem{Kuksa} 
%
V. I.~ Kuksa, Phys. Part. Nucl. {\bf 45}, 998 (2014) (in Russian) 
and references therein.

\bibitem{LFtrans1}
%
K. A. Bugaev,
{Acta. Phys. Polon.} {\bf B 36},   3083-3094 (2005).

\bibitem{LFtrans2}
%
K. A. Bugaev, L. Phair and J. B. Elliott,
{Phys. Rev.} {\bf E 72},  047106  (2005).

{
\bibitem{Rozynek1}
%
J. Rozynek,  J. Phys. G {\bf 42}, 045109 (2015).

\bibitem{Rozynek2}
%
J. Rozynek, Int. J. Mod. Phys. E  {\bf 27}, 1850030 (2018) and references therein.
}

\bibitem{GuggenheimEOS}
%
E. A. Guggenheim,   
Mol. Phys. {\bf 9}, 199  (1965).

{
\bibitem{Bugaev:BEC20}
%
K.~A.~Bugaev, O.~I.~Ivanytskyi, B.~E.~Grinyuk and I.~P.~Yakimenko,
Ukr. J. Phys.  {\bf  65},  963-972 (2020); 
[arXiv:2009.02178 [nucl-th]] 
}


\end{thebibliography}
\end{document}